\documentclass[12pt,preprint]{aastex}
\usepackage{epsfig, natbib}

\shorttitle{H$_2$CO in the Extreme Carbon Star IRC+10216}
\shortauthors{Ford et al.}

\begin{document}

\title{Detection of Formaldehyde Towards the Extreme Carbon Star IRC+10216}
 
\author{K. E. Saavik Ford}
\affil{Department of 
Terrestrial Magnetism, Carnegie Institution of Washington, 5241 
Broad Branch Road, NW, Washington, DC 20015 and Dept. of 
Physics \& Astronomy, The Johns Hopkins University, 
3400 N. Charles St., Baltimore, MD 21218-2686 }
\email{saavik@dtm.ciw.edu}

\author{David A. Neufeld,}
\affil{Dept. of Physics \& Astronomy, The Johns Hopkins University, 
3400 N. Charles St., Baltimore, MD 21218-2686}
\email{neufeld@pha.jhu.edu}

\author{Peter Schilke,}
\affil{Max-Planck-Institut f\"ur Radioastronomie, Auf dem H\"ugel 69, 
53121 Bonn, Germany}
\email{schilke@mpifr-bonn.mpg.de}

\and

\author{Gary J. Melnick}
\affil{Harvard-Smithsonian Center for
Astrophysics, 60 Garden Street, Cambridge, MA 02138}
\email{gmelnick@cfa.harvard.edu}

\begin{abstract}

We report the detection
of H$_2$CO (formaldehyde) around
the carbon-rich AGB star, IRC+10216. 
We find a fractional abundance with respect to molecular hydrogen of
$x({\rm H_2CO})= 1.3^{+1.5}_{-0.8} \times 10^{-8}$.
This corresponds to a formaldehyde abundance with respect to
water vapor of $x({\rm H_2CO})/x({\rm H_2O})=(1.1\pm0.2)\times 10^{-2}$, in 
line with the formaldehyde abundances found in Solar System comets,
and indicates that the putative extrasolar cometary system around
IRC+10216 may have a similar chemical composition to Solar System comets.
However, we also failed to detect CH$_3$OH (methanol) around IRC+10216
and our upper limit of 
$x({\rm CH_3OH})/x({\rm H_2O})<7.7\times 10^{-4}\, (3\sigma)$
indicates that methanol is substantially underabundant in 
IRC+10216, compared to Solar System comets.

We also conclude, based on offset observations, that formaldehyde
has an extended source in the envelope of IRC+10216 and may
be produced by the photodissociation of a parent molecule, similar
to the production mechanism for formaldehyde in Solar System comet
comae.  Preliminary mapping observations also indicate 
a possible asymmetry in the
spatial distribution
of formaldehyde around IRC+10216, but higher signal-to-noise 
observations are
required to confirm this finding.

By serendipity, our observations have led to the 
detection of the $J = 17 - 16$ transition of Al$^{37}$Cl at 
$241.855\,$GHz.  Our analysis of the measured line flux, 
along with those of previously-observed lower 
frequency transitions, 
yields a total AlCl (aluminum monochloride) abundance in the range 
$2 - 8 \times 10^{-8}$ relative to H$_2$; this range, which is a 
factor of 10 smaller than an abundance estimate that has appeared 
previously in the literature, amounts to $\sim 4 - 16\%$ of the 
solar elemental abundance of chlorine, a fraction that is in 
accord with the predictions of thermochemical equilibrium 
models for cool stellar photospheres.

This study is based on observations 
carried out with the IRAM 30m telescope. IRAM is
supported by INSU/CNRS (France), MPG (Germany) and IGN (Spain).

\end{abstract}


\section{Introduction}
IRC+10216 is a nearby well-studied, late-stage carbon-rich Asymptotic
Giant Branch (AGB) star.  It is losing mass at an extremely
rapid rate 
($\dot M \sim 3\times 10^{-5}\, M_{\odot}\,{\rm yr^{-1}}$) \citep{gla96}
and thus possesses a dense, chemically-rich 
circumstellar envelope, shielded from
the interstellar ultraviolet (ISUV) field.
Until recently, only a few well understood
oxygen-bearing molecules had been found around IRC+10216 (CO, and
small amounts of SiO and HCO$^+$), as expected for an 
extremely carbon-rich star 
(C/O$\gtrsim$ 1.4) \citep{gla96}.  
The unexpected detections of water vapor
and OH around IRC+10216 by \citet{mel01} and \citet{for03}, respectively,
were interpreted as evidence for the existence of an
extrasolar cometary system, analogous to the Solar System's Kuiper
Belt, in orbit around IRC+10216.  In this system, the luminosity of
the central star has increased dramatically due to the 
later stages of post-main sequence evolution, causing the icy bodies
in a Kuiper Belt analog to vaporize \citep{SSB90,FN01} and produce
the water vapor observed by \citet{mel01}.  The water vapor
is entrained in the circumstellar outflow, shielded by 
dust, until the dust becomes diffuse enough that water vapor
is exposed to the ISUV.
The water vapor is then photodissociated,
producing the observed OH \citep{for03}.  

The detection of water vapor and OH around IRC+10216 poses the 
interesting question of what
other oxygen-bearing species might be present in the 
circumstellar envelope of IRC+10216.  Assuming that the
presence of water vapor and OH indicates the presence of an
extrasolar cometary system, we should
consider looking for oxygen-bearing species found in
Solar System comets.
The most abundant oxygen-bearing molecules in Solar System
comets are listed in Table \ref{oxymolecules} along with their typical
abundances.  Of the species listed in Table \ref{oxymolecules}, the most 
suitable observational targets in IRC+10216 
are formaldehyde (H$_2$CO) and methanol\footnote{\citet{LC96} 
reported the detection
of millimeter line emission at the frequencies of several methanol
transitions, but subsequently favored C$_4$H and C$_4$H$_2$ as the
species responsible for the emission originally attributed to 
methanol.} (CH$_3$OH).  
Carbon monoxide, while quite
abundant in comets, is expected in any case in large 
quantities in the stellar
outflow \citep{gla96};
thus, the additional CO produced from the vaporization of
icy bodies would not generate a distinctive signature for
an extrasolar cometary system.  
Carbon dioxide, also quite abundant in comets, is
an unsuitable observational target at the present time,
as the atmosphere is opaque at the infrared frequencies of
CO$_2$ vibrational transitions, and there are 
presently no satellites capable of observing in the 
appropriate frequency range.  This leaves us with the
two next most abundant molecules, formaldehyde and methanol,
both of which are observable with the Institut de 
Radioastronomie Millim\'etrique (IRAM) 
$30$ meter ($30$m) radio telescope in Pico Veleta, Spain.
If IRC+10216 contains
similar proportions of formaldehyde and methanol (relative to
water vapor) as Solar System comets, this would
support our hypothesis of an extrasolar cometary system around
IRC+10216.  If, on the other hand, the cometary system around IRC+10216 has a
substantially different chemical composition from Solar System comets,
this could indicate that the cloud that formed IRC+10216 had a
substantially different chemical composition from the protosolar nebula.
A failure to detect expected cometary volatiles like formaldehyde and
methanol around IRC+10216 at abundances roughly consistent with Solar
System comets could also indicate that there is no cometary system around
IRC+10216.

The suggested existence of extrasolar cometary systems is, of course,
not new.  The cometary system around the young star $\beta$ Pictoris 
\citep[see][for an excellent review]{VLF98} has been known and
accepted for more than a decade.  Other similar systems
have been detected \citep[see especially][for the
detection of an extrasolar cometary system around 51 Oph]{rob02}
or suggested \citep[see][for discussions of plausible but unconfirmed
detections]{beu96,BKL01},
so the extrasolar cometary system around IRC+10216
is not the first to be discovered.
But it is the first extrasolar cometary system that can be at least 
partially chemically characterized.  

Before we can understand the chemical composition of
the extrasolar cometary system around IRC+10216, however, we
should examine the chemistry of Solar System comets.
Formaldehyde, specifically, presents some complications.
Formaldehyde was first observed in a comet by \citet{sag86}.
However, it was not until several years later that \citet{mei93}
discovered that formaldehyde is not usually a parent molecule; that is,
formaldehyde is not usually emitted in significant quantities directly from 
comet nuclei.  Instead, it is now thought that 
formaldehyde is 
produced in most comet comae by the photodissociation of a parent molecule
or molecules, 
probably polyoxymethylenes \citep{mei93,cot01,cot04}.  
Recent observations indicate that formaldehyde
is a parent molecule in some Solar System comets 
(Michael A. DiSanti, personal communication);
however, most comets appear to have substantial extended sources
of formaldehyde (indicating that it is primarily a daughter product).
The abundance of formaldehyde in comets, as 
determined from line strengths,
depends on whether the molecule is assumed to have a spatial 
distribution which corresponds to
a parent or a daughter product.
Assuming a daughter 
distribution, as now seems likely for the majority of cases, 
formaldehyde abundances in 
comets are generally found to be a few tenths to a few percent, 
relative to water \citep{boc96}.
An extended distribution of H$_2$CO around IRC+10216 
could indicate that formaldehyde is produced by the
photodissociation of a parent molecule vaporized from
cometary nuclei.
We will need to 
construct a photodissociation model to determine
the expected spatial distribution of formaldehyde around IRC+10216
for the cases of both parent and daughter production,
and carefully compare our models to our observations.
Methanol, by contrast, is a parent molecule in Solar System comets, so our
methanol observations should be easier to interpret.

We explain our observational search for formaldehyde
and methanol below in \S 2.  In \S 3 we
discuss our interpretations of the data.  A summary of our conclusions
is given in \S 4.

\section{Observations and Results}
The main observations discussed in this paper were carried out
using the IRAM 
$30$m radio telescope in late August
and early September of 2002.  For our initial observations, we used the
C150 and D150 receivers to observe the $2_{12}-1_{11}\>$ and  $2_{11}-1_{10}\>$
transitions of formaldehyde, with rest frequencies of $140839.5020$ and 
$150498.3340\,$MHz, respectively.  We also
used the C270 and D270 receivers to observe frequencies around
$241802\,$MHz, which include several transitions of methanol.  
Observations were made in 
wobbler-switching mode, using a wobbling period of $0.5\,$Hz and
a beam throw of $\pm 130\arcsec$.  The telescope pointing and focus
were checked periodically throughout the observing run using
the automated routines {\bf point} and {\bf focus} on Mars, 
Jupiter and Venus.  On each receiver, we used one
$256\,$channel $1\,$MHz filterbank in parallel with the Vespa 
correlator system, set to a resolution of $320\,$kHz and a bandwidth
of $160\,$MHz (for the formaldehyde bands) or $320\,$MHz (for
the methanol band).  For all observations, lines were placed in 
the lower sideband (LSB) with the upper sideband (USB) suppressed
at the $95-97\%$ level.  Because the USB suppression is imperfect, we
also took spectra of the image sidebands and all spectra were observed
with at least two different local oscillator settings, with central
frequencies chosen to avoid prominent lines in the USB.  We are confident
that our spectra contain no contaminating features from the USB.

Most initial data reduction was accomplished by the automated
routines developed by IRAM.
Data from the IRAM 30m telescope is presented to observers
calibrated and in units of antenna temperature, ${\rm T_A^*}$, corrected for
all atmospheric losses.  To convert from units of ${\rm T_A^*}$ to units of
flux, we can simply divide by the gain, which is 
$0.13\,{\rm K\, Jy^{-1}}$ in the $2\,$mm band and 
$0.10\,{\rm K\, Jy^{-1}}$ in the $1.3\,$mm band.
Further data reduction was accomplished using the Continuum and Line
Analysis Single-Dish Software (CLASS) package.
We made two distinct types of observations: 1) integration
centered on 
$\alpha(2000)=9^h47^m57^s.4, 
\delta(2000)=13^{\circ}16^{\prime}44^{\prime\prime}$ 
(the location of IRC+10216) 
and 2) offset mapping observations.
The goal of the first type of observation was simply to detect
formaldehyde or methanol, while the mapping observations were
undertaken after we detected formaldehyde and were aimed at
determining the spatial distribution 
of the formaldehyde emission.  Integration
times for each spectrum are listed in Table \ref{inttime}.  
The analysis
of each set of observations proceeded differently, so
we discuss the analyses separately.

\subsection{Central position observations}
For the central 
position, all observations at a particular frequency 
were summed to produce a total spectrum
for each band, and a linear baseline was subtracted.
These reduced spectra are shown in Figure \ref{cspec}.
The spectra shown from the C150 and D150 receivers are from the
$1\,$MHz filterbank, while the $241.8\,$GHz spectrum comes
from the Vespa correlator system and has been smoothed
once using the CLASS {\bf smooth} command, in order to improve
our signal to noise ratio.
All displayed line fits were produced using the {\bf shell} method 
of line fitting in CLASS.  Line fits produced with the {\bf shell}
routine have four free parameters: area, line center frequency,
expansion velocity (which is a proxy for line width) and
horn-to-center ratio (which characterizes the line shape).
All three bands display some emission features, and we have
identified the carriers of most of these features (see labels
in Figure \ref{cspec} and below).  We have also measured
the line strengths of all features (listed in Table \ref{linefit}).
We discuss each band individually below.

The $241.802\,$GHz band is the simplest, as
there is only one line present.  The line is the $J=17-16$
transition of an isotopologue of aluminum monochloride, Al$^{37}$Cl.
The $J=12-11, 11-10$ and $10-9$ transitions of this 
molecule were previously observed in IRC+10216 
\citep{cer87,CGK00}, but this
is the first observation of the $J=17-16$ transition.
The vertical dotted lines in the spectrum indicate
the position of the brightest expected methanol line
in our spectrum.  There is clearly no evidence for any emission
feature at this frequency.

The $150.498\,$GHz band is more complicated than
the $241.802\,$GHz band, but still it is relatively
simple, as it contains only 
three, well-separated lines.  We identify the
line centered around $v_{LSR}=-26.5{\rm \,km\,s^{-1}}$ as the 
$2_{11}-1_{10}\>$ transition
of formaldehyde at $150.498\,$GHz.
We fit this line with all four parameters 
(area, line center frequency, expansion velocity and horn/center ratio)
free to vary, and find that
$v_{exp}=13.9\pm0.2{\rm \,km\,s^{-1}}$ for this line.  This value is 
somewhat smaller than, but still consistent with, 
the literature value for IRC+10216 of $v_{exp}=14.5$.
The line at roughly
$150.437\,$GHz is identified as the $2_{20}-1_{11}$ 
transition of c-C$_3$H$_2$, 
and the line near $150.386\,$GHz is unidentified
-- we believe the carrier of that feature is not yet 
fully catalogued or perhaps still undiscovered.  \footnote{A deep line
survey of IRC+10216 by Barry Turner (private communication)
confirms our detection of this U-line.  B. T. failed to
detect this line in any of the six
other sources he surveyed: Sgr B2(M), Sgr B2(N), Ori-S, W51M,
W3(IRS5) and G34.3+0.1.  Finally, we note that B. T. attempted
to identify the carrier of this line by constructing line catalogs
for over 2000 species for
which sufficiently precise molecular constants are known, and
which appear possibly relevant to astrophysics.  No match was
obtained, and this failure may indicate that the carrier species
is either so exotic it is not known terrestrially or at least it
has not been properly cataloged.}  In
later discussions we will refer to this unidentified line as U150.  
We note that the fit
for U150 is somewhat affected by a systematic uncertainty
stemming from a number of bad channels in
our filterbank which appear in the blueshifted wing of that line.

The band around $140.840\,$GHz is the most complex.
In it, we see a blend of lines
centered around $v_{LSR}=-26.5{\rm \,km\,s^{-1}}$, the systemic velocity of
IRC+10216.  By eye, we identify one line centered at approximately
$140.853\,$GHz, a second line centered at approximately $140.840\,$GHz
and a third line which has a poorly determined shape and central
frequency but which appears in the redshifted wing of the $140.840\,$GHz 
line.  We identify the line centered at 
approximately $140.853\,$GHz as the
$9_{37}-8_{36}$ transition of NaCN and the line at roughly $140.840\,$GHz
as the $2_{12}-1_{11}\>$ transition of formaldehyde.  The feature in the
redshifted wing of the H$_2$CO $2_{12}-1_{11}\>$ line is
actually two lines; it is the $^2\Pi_{1/2}\>J=59/2-57/2$ doublet of C$_5$H.
The entire blend was
originally identified by \citet{CGK00} as the 
$9_{37}-8_{36}$ transition of NaCN.
Due to the line blending, it was not possible to obtain
a reasonable fit for any of the four lines if all fit parameters
were allowed to vary.  It was also not possible to fit any single
line independently, so for this blend we fit all four lines simultaneously.
In the simultaneous fit, we fixed
the central frequencies and expansion velocities 
($v_{exp}=14.5{\rm \,km\,s^{-1}}$) of the
formaldehyde, C$_5$H and NaCN lines, while allowing the
area and horn/center ratio (the line shape) 
parameters to vary, though we constrained the two C$_5$H lines to have
identical strengths and shapes.  
The four line simultaneous fit is displayed in the top panel of
Figure \ref{cspec}.  We note that NaCN and C$_5$H are
among the list of well studied molecules in IRC+10216 and that the line
fluxes measured here are in line with those measured previously for
transitions of similar excitation \citep[e.g.][]{CGK00}.  As a sanity check, 
we also examined a subtracted
spectrum, removing the fitted C$_5$H and NaCN profiles, presumably
leaving only the $2_{12}-1_{11}\>$ transition of formaldehyde.  This spectrum
is also plotted in Figure \ref{cspec}, and the formaldehyde line is
readily apparent.
The $140.840\,$GHz band
also contains two lines between $140.9$ and $140.95\,$GHz; we identify
these lines as the $9_{36}-8_{35}$ transition of NaCN and
the $6_{25}-5_{24}$ transition of SiC$_2$ 
based on the observations of \citet{CGK00}, 
who observed this
region of the spectrum around IRC+10216 in a previous, shallower 
line survey.  These lines are labelled in Figure \ref{cspec}.
The NaCN line in this group also required a constrained fit due to
the proximity of the strong SiC$_2$ line, as noted in
Table \ref{linefit}.  Table \ref{linefit} also contains fit parameters
for a $^{30}$SiC$_2$ line which we only observed at one local oscillator
setting (due to band edge placement).  We do not plot this
line in Figure \ref{cspec}, 
but it would appear just beyond the blue edge of the
$140.840\,$GHz band plot.

\subsection{Mapping observations}
In addition to observations of the central position, we also
took data at $8$ separate off-center positions, using IRAM's {\bf offset} mode.
All of these observations were radially offset from the center
position by $17\arcsec$, which is a full beam offset at $140\,$GHz
(i.e. the center of the central position is separated
from the center of the offset position by one FWHM of the telescope beam).  
We chose the $8$ position angles such that 
we had two mapping grids or crosses, as drawn in 
Figure \ref{pattern}; one cross
was oriented along the line at position angle $120\,$degrees,
which has been suggested as the plane of a
disk around IRC+10216 \citep{tut00}, while the other cross was
oriented at a position angle of $45\,$degrees to the first cross.
The disk suggested by the observations of \citet{tut00} is only
$\sim 0.05\arcsec$ in radius, however, and thus much smaller than
the scale of our map.
The data are plotted by position angle in Figure \ref{8aspec} for
the $140.840\,$GHz band and Figure \ref{8bspec} for the $150.498\,$GHz band.  
Each plot is a sum of all the data at the labeled position angle
with a first order polynomial baseline removed.
Although we obtained off-center data in the $241.802\,$GHz band, we do not
plot the individual positions here as
the only line in the spectra (Al$^{37}$Cl) is so weak 
that it is not detectable at the individual off-center positions.
Due to the low signal to noise ratio in most of the single
position offset data, we were unable to utilize the
{\bf shell} method of line fitting in CLASS to obtain
consistent fits.  Instead, to determine off-center line strengths 
we integrated the (temperature) value of
each channel in the velocity range $v_{LSR}=-40.1\>$ to 
$-12.9{\rm \,km\,s^{-1}}$, i.e.
$13.6{\rm \,km\,s^{-1}}$ to either side of the systemic velocity of IRC+10216.
We have chosen an expansion velocity,
$v_{exp}=13.6{\rm \,km\,s^{-1}}$, 
because this is the fitted $v_{exp}$ of the bright
SiC$_2$ line in the offset observations.  The smaller
observed expansion velocity (relative to the usual 
$v_{exp}=14.5{\rm \,km\,s^{-1}}$
for this source) is expected for off-center observations, since only 
material at zero offset will be travelling 
directly towards or away from an observer, 
so only that material will appear at $\pm 14.5{\rm \,km\,s^{-1}}$, relative
to the systemic velocity of the source.  Observations
at a full-beam offset therefore do not sample material at the most extreme
velocities.
The formal error on our line strength measurement is just the
r.m.s. noise
(determined from a region of the spectrum with no apparent lines)
times the velocity width 
($27.2{\rm \,km\,s^{-1}}$) of the line, divided by the square root
of the number of
channels in the line, assuming uncorrelated channels.  
We list the line strengths and errors ($1\sigma$) for
both formaldehyde lines and two ``control'' lines, SiC$_2$
$6_{25}-5_{24}$ from the $140.840\,$GHz band and c-C$_3$H$_2$
$2_{20}-1_{11}$ from the $150.498\,$GHz band, 
in Table \ref{offline}.  The line strengths and
errors for the control lines were determined in the same manner as
for the formaldehyde lines.  We refer to the SiC$_2$ and c-C$_3$H$_2$
lines as control lines because we expect that they will have nothing
to do with any cometary system around IRC+10216 (since their presence
in carbon-rich circumstellar envelopes is expected from chemical
models).  The positional variation in the strength of these lines
therefore serves as an indicator of any systematic variation in
line strength with position, either due to instrumental effects, or
due to real variations in the envelope which have nothing to do
with the cometary system.
We note that the method used here for determining line
strengths does not allow for the removal of blended lines, so the
$140.840\,$GHz line strengths at individual offset positions 
contain substantial systematic errors
due to blending.

Since the strength of the formaldehyde lines varies somewhat
with position angle, and the signal to noise ratio at individual
off-center positions is low,
we also created an ``average off-center position'' by summing
over all positions at a particular frequency.  The total off-center
data is plotted in Figure \ref{offspec}, including
the $241.802\,$GHz band.  We determined the 
line strengths of all identified lines at this
off-center position
in essentially the same manner as we did for the central position,
using the {\bf shell} method of line fitting in CLASS.
The results of these fits are listed in Table \ref{offlratio}.

\subsection{Followup Observations}
We also conducted followup observations in early
2004 to detect additional lines of formaldehyde, 
also using the IRAM $30$m.  We used the A230 and B230 receivers to observe
the $3_{13}-2_{12}$, $3_{03}-2_{02}$ and $3_{12}-2_{11}$ transitions of 
formaldehyde, with rest frequencies
of $211211.468$, $218222.192$, and $225697.775\,$MHz, respectively.  
We also used an experimental setup
of the B100 receiver in an unsuccessful attempt to detect the 
$1_{01}-0_{00}$ transition of formaldehyde
at a rest frequency of $72837.948$MHz.  We again 
used the wobbler switching mode, with the same
period and beam throw, and we used the same calibration procedures as
for our initial observations.  On each receiver, we used one
$256\,$channel $1\,$MHz filterbank in parallel with the Vespa 
correlator system, set to a resolution of $320\,$kHz and a bandwidth
of $160\,$MHz (100 GHz receiver) or $320\,$MHz (for
the 230 GHz receivers).  For all of our observations, lines were placed in 
the lower sideband (LSB).  Data reduction was the same
as for our initial observations, except as noted in discussions below.
Our followup observation spectra are plotted in Figure \ref{folspec} with
line fits; we discuss each spectrum separately.

Our $72.838\,$GHz observations were conducted using a new receiver
setup which involved tuning the B100 receiver to frequencies below
the nominal limits of the system, and using the LO system from the B230
receiver.  The spectrum plotted is from the $1\,$MHz filterbank.  
Owing to operational difficulties with the system, we obtained data
at only one LO setting and we did not obtain a spectrum of the image band.  
The system was still undergoing commissioning tests during
our observations and flux calibrations may not be entirely reliable.
The single observed line in this spectrum, the HC$_3$N $J=8-7$ line,
has been observed only once before, by \citet{mor75} and our observation
is consistent with their flux measurements, but owing to the relatively large
flux uncertainties of the \citet{mor75} observation, 
we are unable to obtain a better
calibration, and estimate that our systematic uncertainty may be as high
as $50\%$.  The vertical dashed lines in this spectrum 
indicate the expected position of the
$1_{01}-0_{00}$ transition of formaldehyde.

The spectrum plotted around $211.211\,$GHz represents data taken at
two LO settings and is from the $1\,$MHz filterbank.  
The spectrum shows only two features.  We identify
the line at $v_{LSR}=-26.5\,{\rm km\,s^{-1}}$ as
the $3_{13}-2_{12}$ transition of formaldehyde.  The remaining feature
is due to a blend of the $5_{15}-4_{14}$ $11/2-9/2$ $6-5$ and $5-4$ 
transitions of c-C$_3$H.

The $218.222\,$GHz spectrum was obtained using a wider bandwidth than the rest
of our spectra.  This spectrum was obtained using the VESPA correlator
system and stretches from $218.084$ to $218.479\,$GHz (the portion
which is not displayed contains no obvious features).  The spectrum has been
smoothed to a resolution of $1.25\,$MHz per channel to improve our
signal to noise ratio.  This spectrum represents only one LO setting, though
we were able to obtain a spectrum of the image band.  It confirms that
the main features in the displayed spectrum are uncontaminated by features
from the image band.  There are 6 lines in this spectrum, only 3 of which
can be identified.  The largest feature, near $218.32\,$GHz is the
$J=24-23$ transition of HC$_3$N.  The next line, at around 
$218.29\,$GHz is unidentified, and we label it U218a.  There does
appear to be some non-zero flux between U218a and the HC$_3$N line;
however, we are unable to associate it with a carrier or even a rough
line fit so it is unlabelled.  We identify the line near $218.22\,$GHz
as the $3_{03}-2_{02}$ transition of formaldehyde.  The line near
$218.16\,$GHz is the $5_{24}-4_{13}$ transition of c-C$_3$H$_2$.  The
two line blend at the red edge of the spectrum consists of two unidentified
lines, labelled U218b and U218c.  In order to obtain reasonable line
fits it was necessary to simultaneously fit U218b and U218c, and to
constrain the expansion velocity of the fit for U218c.
The $225.697\,$GHz spectrum shows only one line, the
$3_{12}-2_{11}$ line of formaldehyde.

Given that we have carried out a very deep 
integration towards IRC+10216, it is worth considering 
the odds that we have misidentified the putative 
H$_2$CO lines due to confusion with previously 
uncataloged, unidentified lines, or U-lines.  If we take as our 
null hypothesis that
H$_2$CO is not responsible for any of the lines in any of our spectra,
we find that there is one U-line in our $140.8\,$GHz spectrum, two in
our $150.5\,$GHz spectrum, one in our $211.2\,$GHz spectrum,
four in our $218.2\,$GHz spectrum, and one in our 
$225.7\,$GHz spectrum.  We can then 
determine the probability that five of the
nine U-lines should - by random chance - lie at the five H$_2$CO frequencies.
This is the probability that our identification is in error.

Our bandwidth for each of these spectra is $256\,$MHz, except
for the $218.2\,$GHz spectrum where it is $395\,$MHz.
Our null hypothesis is that there are no formaldehyde lines
in our spectra, and that the lines we have been identifying as
formaldehyde are actually U-lines.  The
rate of occurence of U-lines in each spectrum is just the number
of U-lines in that spectrum divided by the number of independent
resolution elements within that band.  For all spectra except the
$218.2\,$GHz spectrum, the resolution elements are $1\,$MHz wide;
for the $218.2\,$GHz spectrum, the resolution 
elements have been smoothed to $1.25\,$MHz wide.
Since any line fit which has
a central frequency within $\pm0.5$ a resolution element
of the laboratory
formaldehyde frequency would likely be identified
as a transition of formaldehyde, the probability that
a single line would meet that criterion is $P=1/256,\>
2/256,\>1/256,\>4/316,$ and $1/256$, respectively for the
$140.8,\>150.5,\>211.2,\>218.2$ and $225.7\,$GHz spectra.
Thus, the odds that we would simultaneously confuse five
independent U-lines with the five observed transitions of formaldehyde is only
$P=(1/256)(2/256)(1/256)(4/316)(1/256)\sim 5.9\times 10^{-12}$.  
Our confidence in this formaldehyde
identification is strengthened by the fact that the ratio
of line fluxes of the 
$2_{11}-1_{10}\>$ to $2_{12}-1_{11}\>$ transition is observed to
be $0.58\pm0.03$, in agreement with the predicted value of 0.56.

\section{Discussion}
\subsection{Models for formaldehyde emission}
We have constructed
several models to predict the strength and spatial
distribution of formaldehyde emission around IRC+10216.
In order to predict the
emission characteristics, we needed to determine both the expected spatial
distribution of formaldehyde and the excitation conditions
of the molecule.  To determine the spatial distribution of formaldehyde, 
we were able to use a photodissociation model similar to that
discussed for H$_2$O and OH 
in \citet{for03}.
We assumed that formaldehyde
could be either a parent molecule (outgassed directly from
an icy nucleus, like H$_2$O) or a daughter product (created when
some larger parent molecule was photodissociated, as with OH).
We assumed that the vaporized molecule (either formaldehyde or
its unidentified parent) is injected into 
the outflow at $100\,$AU \citep[based on calculations by][]{FN01}, 
and that the vaporized molecule 
travels undisturbed 
through the
outflow until it is photodissociated.  If the assumed vaporizing
molecule
is not formaldehyde, the parent is
photodissociated into formaldehyde, and the formaldehyde
is also subsequently photodissociated.
We assume a photodissociation rate of the form
$\Gamma_i (A_v)=2 C_i G_0
\exp(-\alpha_i A_v - \beta_i A_v^2)\, {\rm s^{-1}},$
where $i$ represents either a parent molecule or H$_2$CO, 
$A_v$ is the depth in visual magnitudes and $C_i$, $\alpha_i$ and $\beta_i$
are the relevant photodissociation parameters for a cloud
assumed to be spherical and uniformly illuminated from the
exterior by the ISUV field.
The factor $G_0$ represents the relative strength
of the ISUV field and $G_0=1$ defines the average ISUV field
\citep{dra78}.
Since the hypothetical parent of formaldehyde is unknown, we do not
have tight constraints on the values of 
$C_{p}$, $\alpha_{p}$ and $\beta_{p}$, the photodissociation
parameters of the parent molecule.
Based on the findings
of \citet{mei93},
we initially assumed that
$C_{p}=C_{H_2CO}=7.81\times 10^{-10}\,$s$^{-1}$.  We also
assumed $\alpha_{p}=\alpha_{H_2CO}=2.22$ 
and $\beta_{p}=\beta_{H_2CO}=0.0$, since there are no
observations which place useful
constraints on these parameters.  The values of
the photodissocation parameters for H$_2$CO were determined using
the method described in the appendix of \citet{for03} based on
plane-parallel parameters for H$_2$CO from \citet{rob91}.
Under the assumption that the photodissociation parameters
for the unknown formaldehyde parent molecule 
are identical to the parameters for formaldehyde, we can
run a single model to determine the spatial distribution of
formaldehyde as either a vaporizing molecule or daughter product.  We
plot the results of this calculation in the
upper panel of Figure \ref{photdiss}.
The photodissociation model we used was the standard model, A,
from \citet{for03}, that is, $G_0=1$, 
$A_v/N_H=5.6\times 10^{-22}\,{\rm mag\, cm^{2}}$
and $\dot M = 3 \times 10^{-5}\, {\rm M_{\odot}\, yr^{-1}}$.

Given a spatial distribution for the formaldehyde, we can
calculate the excitation conditions of the molecule and
the line strengths of various transitions.  For these
calculations, we employed the Large Velocity Gradient (LVG)
code of \citet{NK93}.  
The LVG code uses the escape probability
method to solve the equations of statistical equilibrium for
the energy levels of formaldehyde.  We used the same
parameters for the circumstellar conditions as
\citet{mel01}, listed here in Table \ref{CSEparams}.
We obtained the formaldehyde Einstein-A coefficients and 
the energy levels from \citet{jar86}, 
and the collisional coefficients
from \citet{gre91}.  Given an assumed abundance and spatial 
distribution for formaldehyde, the LVG program calculates both the total 
line strength for various transitions, 
and (for each transition)
the amount of line luminosity, $dL$, produced in
a given radial interval, $dr$, as a function of $r$, where $r$ is the
astrocentric radius.  We have a third program
which uses the tabulated $dL(r)/dr$ to calculate the formaldehyde
line strength and shape after convolution
with the telescope beam and Gaussian microturbulence.  
The IRAM 30m telescope has a frequency-dependent $HPBW$ of $17\arcsec$ at
$140\,$GHz (near
the $2_{12}-1_{11}\>$ 
transition of formaldehyde), $16\arcsec$ at $150\,$GHz (near
the $2_{11}-1_{10}\>$
transition of formaldehyde) and $10\arcsec$ around $240\,$GHz.
The convolution program is similar to the line shape
program of \citet{for03}, except that this program
also allows calculations of line strengths at offset positions.
We can compare the output of the convolution program
directly to our observations.

\subsection{Comparison of models with observation}
In comparing our models with our observations,
we will rely primarily on the $150.498\,$GHz $2_{11}-1_{10}\>$
transition of formaldehyde, for reasons we detail
below.
In order to accurately determine the
abundance of formaldehyde around 
IRC+10216, we must
first determine its radial distribution, since
any abundance determination will depend on the excitation
conditions of formaldehyde, which in turn depends on
formaldehyde's radial distribution.  
Fortunately, our offset observations are ideally
suited to distinguishing whether formaldehyde has
a compact or extended source in IRC+10216.  We find,
based on our ``average off-center position'', that the ratio of
$(\int T_A^*dv)_{off}$, the integrated antenna temperature 
at the off-center position, to $(\int T_A^*dv)_{cent}$, the
integrated antenna temperature at the central 
position is $0.40 \pm 0.03$ at the $150.498\,$GHz $2_{11}-1_{10}\>$
transition
and $0.51 \pm 0.02$ at the $140.840\,$GHz
$2_{12}-1_{11}\>$ transition.
Here we are quoting statistical errors only ($1\sigma$), and we point out
that the $2_{12}-1_{11}\>$ transition may be partly contaminated by the
C$_5$H doublet.  The contamination from the C$_5$H 
lines is likely to be higher in the
off-center data, due to the increased relative strength of the doublet in
our off-center spectrum.  We therefore ignore the ratio for the 
$140.840\,$GHz $2_{12}-1_{11}\>$ transition,
as we consider it to be unreliable,
and concentrate on the ratio for the 
$150.498\,$GHz $2_{11}-1_{10}\>$ transition.
We can compare this ratio to the predicted ratio obtained from line strengths
calculated by our
convolution program.  
We find, for widely varying formaldehyde abundances,
$(\int T_A^*dv)_{off}/(\int T_A^*dv)_{cent}=0.09$ 
if formaldehyde is a parent molecule,
or $(\int T_A^*dv)_{off}/(\int T_A^*dv)_{cent}=0.3$ 
if formaldehyde is a daughter molecule
whose parent has identical photodissociation parameters to
formaldehyde.  We can obtain
a best fit ratio, $(\int T_A^*dv)_{off}/(\int T_A^*dv)_{cent}=0.4$ 
for $C_{p}=1.6\times 10^{-10}\,$s$^{-1}$
(we still assume $\alpha_{p}=2.22$ and $\beta_{p}=0.0$, since
our observations are not well suited to determining these parameters
and these values for $\alpha$ and $\beta$ are fairly typical
for a variety of molecules characterized by \citet{rob91}).  
Clearly, formaldehyde
has an extended source in the envelope of IRC+10216 and a
parent molecule with the same photodissociation parameters
as formaldehyde is not quite permitted by the data.
We note that $C_{p}$ is not well constrained by our observations,
as large variations in $C_{p}$ produce smaller variations in
the ratio $(\int T_A^*dv)_{off}/(\int T_A^*dv)_{cent}$.  We plot
$C_{p}$ versus $(\int T_A^*dv)_{off}/(\int T_A^*dv)_{cent}$ in
Figure \ref{ratio}.
We also plot the spatial
distribution of formaldehyde for our best fit $C_{p}$ 
in the bottom panel of Figure \ref{photdiss}.

Based on our knowledge that formaldehyde has an extended source
in the envelope around IRC+10216, and using the best fit
photodissociation parameters for formaldehyde's parent molecule, 
we can now determine the peak
abundance of formaldehyde, relative to molecular hydrogen, which we find to be 
$x({\rm H_2CO})=1.3^{+1.5}_{-0.8}\times 10^{-8}$, assuming
an ortho-to-para ratio of 3:1, as expected in LTE.  As with
water vapor, the abundance depends
on the assumed stellar mass loss rate, and higher abundances
correspond to lower mass loss rates.  
This abundance determination is based on
the $2_{11}-1_{10}\>$ transition line strength 
only, as line blending produces systematic
errors in the measured strength of the $2_{12}-1_{11}\>$ transition
which are
difficult to estimate; however, our models predict
a line strength for the $2_{12}-1_{11}\>$ transition of 
$(\int T_A^*dv)_{2_{12}-1_{11}}=701\pm37\,
{\rm mK}{\rm \,km\,s^{-1}},\>(1\sigma)$ for
the abundance range quoted above.  These predicted fluxes are quite
consistent with the measured value of 
$(\int T_A^*dv)_{2_{12}-1_{11}}=687\pm19\,
{\rm mK}{\rm \,km\,s^{-1}},\>(1\sigma)$.
In addition, the measured relative strengths of the formaldehyde lines are in
good agreement with the relative strengths predicted by our LVG
excitation code.

In contrast with our 2mm observations, the lines 
detected in our followup observations are stronger than predicted
by factors of $\sim 3-8$.  We believe these discrepancies can
be attributed to the molecular collision rate coefficients we used
in our LVG code,
specifically those computed by \citet{gre91}.  These collision
rates were computed for collisions of formaldehyde with
atomic helium, while in the circumstellar envelope of IRC+10216
(and in most other astrophysical instances) the dominant 
collision partner will be molecular hydrogen.  The use
of atomic helium as a substitute partner in collision models is
nevertheless routine, due to the fact that the potential of
atomic helium is vastly simpler to deal with computationally
than the potential of molecular hydrogen, thus enabling the calculation
of collision rates for a fairly large selection of molecules in
a reasonable amount of time.  As suggested by \citet{gre91}, in order to 
use the H$_2$CO-He collision rates for H$_2$CO-H$_2$ collisions we applied a
uniform correction factor ($2.2$) to the H$_2$CO-He collision rates to
account for the difference in the cross-sections and relative 
collision velocities of He and H$_2$.
However, \citet{gre91} also notes there is likely some variation in
the detailed state to state collision rates owing to the detailed
differences between the He and H$_2$ potentials.  Specifically,
ammonia is one of the few molecules for which both He and H$_2$
collision rates have been calculated and comparisons with the findings
of those studies are instructive.  In the case of ammonia collisions,
it was found that for transitions where $\Delta J > 1$, rotationally
excited H$_2$ caused transition rates to be enhanced by a factor of
$2-3$.  Additionally, even for excitation by H$_2(j=0)$ rates were
usually only within a factor of 2 of the rates for excitation by
He, though some were larger and some were smaller, with no obvious pattern.
The analysis of our observations is thus severely impeded by the lack
of reliable collision rates; however, since the $J=3-2$ transitions
(i.e. all of the 1.3mm line detections) would be preferentially affected by
enhancements of collision rates for $\Delta J > 1$ transitions,
we regard these lines as less reliable indicators of the formaldehyde
abundance in IRC+10216 and we do not rely on them for our
abundance estimate.  Given the uncertainties in all of our
collision rates, we must be cautious in interpreting our calculated
abundances.  While we consider the 2mm transitions to be the {\it most}
reliable abundance indicators, even they may lead to abundance estimates
which are off by factors of $2$ or possibly more.

Another discrepancy between our predictions and observations is
posed by our non-detection of formaldehyde at
$72.838\,$GHz, where the $3\sigma$ upper limit is 
weaker than the line strength predicted by our standard model 
by a factor of about $1.8$.  While collision rates may play a role here,
we believe the overprediction can be primarily attributed to two effects:
1) the uncertain calibration of observations at this frequency (possibly
up to $\sim 50\% $ and 2) the sensitivity of this transition to
the size of the emitting region.  Due to the low frequency of these
observations, the $72.838\,$GHz line was subject to the largest
beamsize of any of our observations.  If the UV field was slightly
stronger than we assumed, or the unshielded photodissociation rate of
formaldehyde (which we estimate to be accurate to 
no better than $20\%$) was slightly larger than we assumed, the size of the
formaldehyde emitting region would shrink somewhat, and the predicted flux at
$72.838\,$GHz would drop, while the flux for higher frequency lines would
remain unaffected.
Since we were unable to detect the $1_{01}-0_{00}$ transition at $72.838\,$GHz,
our only detected transition of
para-H$_2$CO is the $3_{03}-2_{02}$ transition;
in the absence of more
reliable collision rates we believe we cannot accurately model the
excitation of this transition and cannot rely on it to
determine the ortho:para
ratio of H$_2$CO in this source.  Since our models of the
$1_{01}-0_{00}$ transition are likely the more reliable of the
two para-H$_2$CO lines, we find that our observations are not
inconsistent with the LTE ortho:para ratio of 3:1, which we therefore assume
throughout this paper.

Despite the uncertainties, our 
determined abundance agrees well with measured
values of the formaldehyde abundance relative to water
vapor in Solar System comets.  If we take 
$x({\rm H_2O})=(4-27) \times 10^{-7}$ around IRC+10216,
our peak formaldehyde abundance with respect to water is
$x({\rm H_2CO})/x({\rm H_2O})=(1.1\pm0.2)\times 10^{-2}$.  
The fractional error on the relative abundance of formaldehyde
to water vapor is much smaller
than the error on either the water vapor abundance or the
formaldehyde abundance because the primary source of error
in the absolute abundance measurements is the stellar mass loss
rate, and this affects both absolute abundances similarly,
so the error tends to cancel out, allowing us to determine
the relative abundance very precisely.
This compares well
with the peak values measured for Comet Lee
($x({\rm H_2CO})/x({\rm H_2O})=
1\times 10^{-2}$) \citep{biv00}, Comet Hyakutake 
($x({\rm H_2CO})/x({\rm H_2O})=8\times 10^{-3}$) \citep{biv99}
and several
comets observed by \citet{col92}
($x({\rm H_2CO})/x({\rm H_2O})\sim5\times 10^{-3}$).

While we do not know the identity
of the carrier of U150, we
can also use our photodissociation model to make
some useful statements about possible carriers of the
U150 line.
The carrier of the U150 line is likely to
be a product of the rich photochemistry in the circumstellar
envelope of IRC+10216.  The line has a high off-center/central-position
ratio, comparable to or greater than the off-center/central-position
ratio of the formaldehyde line in the $150\,$GHz band,
indicating that the carrier is at least as extended as formaldehyde.
We can obtain additional information about the spatial
distribution of the U150 line carrier by examining
its line shape in our central and off-center spectra.
The U150 line has a double-peaked profile in our
central position observations, indicating a resolved source, i.e.
the emission is much larger than the telescope beam ($HPBW=16\arcsec$).  
By contrast, the off-center line shape for U150 is relatively
flat-topped, particularly compared to the c-C$_3$H$_2$ line, 
as would be expected for a source which is
less extended than the c-C$_3$H$_2$ line.  This conclusion is
supported by the off-center/central-position ratio for the
c-C$_3$H$_2$ line, which is larger than for the U150 line.
In addition, the existence of relatively bright emission in U150 
so far from IRC+10216
indicates that the line originates in a relatively
low-lying state which is easily excited even at
the lower densities found beyond 
$\sim1400\,$AU from the star.

\subsection{Chemical production of formaldehyde?}
A recent paper by \citet{wil04} has suggested that the water
vapor and OH found around IRC+10216 could be formed chemically by
Fischer-Tropsch catalysis on grain surfaces.  While 
this mechanism is incapable of directly producing large
amounts of formaldehyde (comparable to what we see in IRC+10216),
Willacy suggests that the presence of formaldehyde could be
explained separately by the photodissociation of methanol, which
could be formed in large quantities by Fischer-Tropsch reactions.
Our failure to detect methanol in significant quantities rules
out this method of production of formaldehyde, and also suggests
that grain catalysis is not the source of the 
water vapor and OH around IRC+10216.  Nevertheless, there
are plausible chemical production routes for formaldehyde
which we examine here.  There are at least two important
reactions that are responsible for the production
of formaldehyde around oxygen-rich asymptotic
giant branch stars, which may also operate around
IRC+10216.  These are 
\begin{equation}
{\rm OH} + {\rm CH_2} \to {\rm H_2CO} + {\rm H} 
\end{equation}
and
\begin{equation}
{\rm O} + {\rm CH_3} \to {\rm H_2CO} + {\rm H}.
\end{equation}
In both oxygen-rich
stars and in IRC+10216, the CH$_2$ and CH$_3$
would be produced by the photodissociation of methane
(CH$_4$).  The source of OH around IRC+10216 and
oxygen-rich stars would also be the same, namely
the photodissociation of water vapor.  The atomic
oxygen required for the second reaction would come
almost exclusively from the photodissociation of
CO in the case of IRC+10216, but for oxygen-rich
stars, both the CO and OH will make a contribution.
In both types of stars atomic oxygen will be
available in the outermost regions of the circumstellar
envelope.  Clearly then, the reactants are available around IRC+10216, 
and since they are themselves photodissociation products,
they will result in an extended source for formaldehyde, consistent
with our observations.  We must therefore calculate if chemical
reactions will produce enough formaldehyde to match our
observations. 

A rough estimate of the expected
formaldehyde abundance may be made by comparing
the methane and water vapor abundances around
IRC+10216 to that around oxygen-rich stars.
\citet{WM97} have modeled the chemistry
of several oxygen-rich stars, including the reactions
listed above for creating formaldehyde.  They typically
find, for assumed abundances of $x({\rm H_2O})=3\times 10^{-4}$
and $x({\rm CH_4})=3\times 10^{-5}$,
a peak formaldehyde abundance of $x({\rm H_2CO})\sim 10^{-6}$.
Since the observed $x({\rm H_2O})$ and $x({\rm CH_4})$
for IRC+10216 are down by factors of 
$\sim300$ \citep{mel01} and $\sim10$ \citep{KR93}, 
respectively,
and because $x({\rm H_2CO})$ is dependent on the product
of $x({\rm H_2O})$ and $x({\rm CH_4})$, we expect
$x({\rm H_2CO})$ to be down by a factor of roughly 3000 relative
to the models of \citet{WM97},
or $x({\rm H_2CO})\sim 3\times 10^{-10}$.  This is about
2 orders of magnitude less than is observed in IRC+10216,
which does not make the chemical explanation look very promising.
We can also do a more careful calculation, modeling
the individual chemical reactions which produce
formaldehyde.  Models by \citet{PMM03}
tabulate the abundances of CH$_2$, CH$_3$ and atomic oxygen
as a function of radial distance from IRC+10216, and
we have a model from \citet{for03} which tabulates
the abundance of OH as a function of radius.  If we
assume that the creation of formaldehyde is a perturbation
to these abundances (i.e. the reaction does not consume
large quantities of any of the reactants), we can
simply model the abundance of formaldehyde as a function of radius with
the equation
\begin{equation}
\dot x({\rm H_2CO})=k_1 n_{\rm H_2} x({\rm OH})x({\rm CH_2})
+k_2 n_{\rm H_2} x({\rm O})x({\rm CH_3})-\Gamma_{\rm H_2CO}x({\rm H_2CO}).
\label{eq}
\end{equation}
Here $\dot x({\rm H_2CO})$ is the time derivative of $x({\rm H_2CO})$,
where time and radius are related through the expression
$R=v_{exp}t$, $n_{\rm H_2}$ is the density of molecular hydrogen,
$k_1=3.0 \times 10^{-11}$ and $k_2=1.3 \times 10^{-10}$ are 
the appropriate reaction rates in units of
cm$^3\,$s$^{-1}$ \citep{LMM00} and $\Gamma_{\rm H_2CO}$ 
is the photodissociation
rate of formaldehyde.  Our calculation
is slightly complicated by the fact that the H$_2$O abundance,
and therefore the OH abundance, is somewhat uncertain.  But
we can easily calculate the expected abundance of
formaldehyde with respect to water vapor, 
$x({\rm H_2CO})/x({\rm H_2O})$,
by simply dividing our calculated $x({\rm H_2CO})$ by
our assumed $x({\rm H_2O})$.  Integrating equation \ref{eq},
and assuming $x({\rm H_2O})=1.2\times10^{-6}$,
we find a peak formaldehyde abundance of 
$x({\rm H_2CO})=1.1\times 10^{-9}$,
which is
slightly larger than our rough estimate, but still
an order of magnitude smaller than our observed value.
Importantly, we find that for all plausible values of 
$x({\rm H_2O})$, chemical reactions
can only produce $x({\rm H_2CO})/x({\rm H_2O})\sim0.1\%$, or
about an order of magnitude less than our observed ratio.
We plot the spatial distribution of chemically produced
formaldehyde, as well as the relevant reactants in
Figure \ref{chem}, for $x({\rm H_2O})=1.2\times10^{-6}$.  
We note here that if
we were to treat the problem not as a perturbation to
the chemistry, but were to allow the formaldehyde
reaction to consume reactants, this would further depress
the calculated $x({\rm H_2CO})$.  In view of these calculations,
we believe we can safely rule out any chemical explanation
for the presence of formaldehyde around IRC+10216.

\subsection{Hints of asymmetry?}
The mapping observations we obtained contain tantalizing
hints that the distribution of formaldehyde around IRC+10216 is not
spherically symmetric.
If we look at the off-center data separately by position
angle (as opposed to the ``average off-center position'' we discussed
above), we see that a positive signal was detected 
at both $140.840$ and $150.498\,$GHz at all positions, though several
of the positions have signals that are not statistically significant
(see Table \ref{offline}).
We have plotted the data from Table \ref{offline}
for the H$_2$CO $2_{11}-1_{10}$, c-C$_3$H$_2$ $2_{20}-1_{11}$ and 
SiC$_2$ $6_{25}-5_{24}$ lines, with the
line strengths normalized relative to the average line strength
at each frequency, in Figure \ref{position} (we ignore
the H$_2$CO $2_{12}-1_{11}\>$ data due
to sytematic errors from line blending 
which are difficult to eliminate).  
The c-C$_3$H$_2$ $2_{20}-1_{11}$ and 
SiC$_2$ $6_{25}-5_{24}$ lines are our ``control lines'',
since c-C$_3$H$_2$ and SiC$_2$ are molecules that
are produced normally in a carbon-rich circumstellar
envelope.  These molecules should be entirely unrelated to the
vaporizing Kuiper Belt analog.  Additionally, our control
lines were bright, unblended and do not contain any bad channels,
so we believe that the systematic
errors associated with the measured line strengths will
be neglible.
We can see that
the line strengths of the control lines do not vary
much with position angle.  By contrast, the formaldehyde
line varies substantially, in one case, at position angle
$210^{\circ}$, by more than $3\sigma$
from the average value.  Since there are $8$ data points,
the probability of such a large variation occurring by random
chance at any $1$ of the $8$ points
is $4\times 10^{-3}$, but
this still represents an exciting hint that formaldehyde may have
a different spatial distribution than other molecules 
with well-understood formation mechanisms
in the circumstellar envelope of IRC+10216 (based on 
our two control lines).
We expect that formaldehyde would have a spatial distribution
independent of these other ``native'' molecules if the
formaldehyde were produced by
outgassing from a Kuiper Belt analog; in that case, we
would still expect the spatial
distribution of formaldehyde to track that of water vapor and OH.
We have a very poor idea of the spatial distribution of either
the water vapor or the OH, though, as discussed previously, the OH
may show signs of being asymmetrically distributed.  One natural
model for the spatial distribution of all three of 
these molecules is, of course,
a disk or ring structure, which would display departures
from circular symmetry if we observed the ring more or less edge-on.
If we observe an essentially edge-on ring, we would
expect our data to show evidence of periodicity,
with a period of $180^{\circ}$.  There does not appear to be
an elevated signal $180^{\circ}$ away from the
elevated point at $210^{\circ}$, as would have been expected for an edge-on
system; however, at the
present signal to noise level, it is difficult to rule out
a ring with any certainty.

Our tentative hints of asymmetry
in the distribution of formaldehyde are especially interesting in 
light of OH observations by \citet{for03}.
We believe that both the OH and formaldehyde are
photodissociation products of parent molecules which are
released from cometary nuclei in orbit around IRC+10216.
Therefore, we might expect H$_2$CO and OH to have similar spatial
distributions and line shapes.  Instead, OH and H$_2$CO possess
very different line shapes, indicating different spatial distributions.
We do not fully understand why OH displays a narrow, blue-shifted
line profile, and like \citet{for03}, 
we believe that the spatial distribution of OH around IRC+10216
requires further study.  We can think of
no production mechanism for H$_2$CO in the envelope of IRC+10216
that is not closely related to OH; we are therefore 
at a loss to explain the clearly differing line shapes.
In addition, the possible asymmetries
noted in the OH and 
formaldehyde distributions are not the same -- OH has
a front-to-back asymmetry, while formaldehyde's
asymmetry is in the plane of the sky (if it exists at all).
Also, formaldehyde does not appear to be substantially 
asymmetrically distributed in the inner regions of
the outflow around IRC+10216, as evidenced by the
shapes of the formaldehyde lines in our central
position observations.  Unfortunately, we do not
have similar information on the spatial distribution
of OH, since Arecibo has a much larger beam than
IRAM, and we do not have mapping observations of
the OH from Arecibo.  Clearly these discrepancies
require further study -- ideally, deep maps of the
distribution of both formaldehyde and OH.  Further
mapping observations of formaldehyde around IRC+10216
should be not be difficult or enormously time consuming,
and such maps will determine if the hints of formaldehyde's
asymmetry are real.
Mapping of OH around IRC+10216, on the other hand, 
would consume large amounts
of telescope time, due to the faint OH signal (sure to become
fainter for off-center observations).  Still, we view
this as an extremely interesting and important observation,
and it may be possible to complete an OH mapping campaign
without placing undue demands on telescope time
if the campaign were stretched over several observing seasons.

\subsection{Methanol}
If we believe that the formaldehyde around IRC+10216 is
produced by vaporization from icy bodies with chemical compositions
similar to Solar System comets, we would also expect to
detect methanol in substantial quantities.  Methanol
is usually present in Solar System comets at the few percent level
(relative to water vapor) \citep{boc96}.
In individual comets, methanol is usually
observed to be more abundant than formaldehyde 
\citep[see][and references therein]{col92,biv99,biv00,EC00}.  
Ices in Solar System comets, in turn, are believed to come from
the interstellar medium, and are expected to
have a similar chemical composition to
interstellar ices, particularly 
those found in molecular hot cores \citep{EC00}.  Indeed,
both the methanol and formaldehyde \footnote{The precise link 
between formaldehyde in hot cores and formaldehyde in
Solar System comet comae remains unclear -- some processing may occur
between the ISM and comet formation, as formaldehyde is present in comets
as a daughter product, and is not present in cometary ices as a
stand-alone molecule, though it may be present as part of a formaldehyde
polymer, known as a polyoxymethylene \citep{BC02}.}
found in comets is 
believed to form from grain surface reactions in the ISM,
through successive hydrogenation reactions with CO.  Observations
of hot cores indicate that their methanol abundances can vary
widely, with methanol ice abundances as high as $30\%$ relative to 
water ice in some sources, 
and upper limits in other sources as low as $2\%$ \citep{dar99}.
Very small abundances of methanol are difficult to measure
in the ice phase, due to the fact that the spectral features
for solids are broad and difficult to distinguish at low abundance.

Our expected line strengths for methanol were
based on models kindly run by Silvia Leurini (private communication),
using collision rates from \citet{PFD01,PFD02}.
Our observations could have detected
methanol lines as faint as $T_A^*>10\,{\rm mK}\, (3\sigma)$.
Models by S. L. indicate
that the brightest line in our band would be the 
$5_{00}-4_{00}$ transition
of the A species of methanol at $241791.3\,$MHz, and
we could have detected this line of methanol at
abundances as small as $x({\rm CH_3OH})>8.5\times 10^{-10}\,(3\sigma)$ or
$x({\rm CH_3OH})/x({\rm H_2O})<7.7\times 10^{-4}$.
Our non-detection therefore requires some explanation, if
we believe the formaldehyde comes from comet-like bodies.
One possible explanation is that the methanol is
being pumped by the large IR luminosity of IRC+10216 
into vibrationally excited states, and 
the transitions between low-lying states 
we attempted to observe are substantially fainter than 
expected due to depopulation of the low-lying states; however,
a full excitation calculation including IR pumping
is beyond the scope of this paper (and is also not presently
available from S. L.).
Another possible explanation is that there is no
methanol at all, or at least that it is substantially
underabundant, compared with methanol in normal Solar System
comets.  This could simply be the result of different
chemistry in the natal cloud of IRC+10216 (as compared
with our Solar System), which resulted in a depletion
of methanol for Kuiper Belt Object (KBO) 
analogs that formed around IRC+10216.  This is not
entirely implausible, considering the wide range
of observed methanol abundances in star forming
regions (molecular hot cores).
It is also possible that comets around IRC+10216 are
more similar to a special class of comets, which, in our
Solar System are relatively rare.  There exists a 
class of comets which are
methanol-poor ($x({\rm CH_3OH})/x({\rm H_2O})<0.17\%$), 
but which have essentially normal
formaldehyde abundances \citep{mum02}.  It is not
understood how these comets became methanol-depleted
in our own Solar System, so we are truly at a loss
to explain why {\it most} comets in another system
would be so methanol-poor.  
Further study of methanol-poor comets in the
Solar System is clearly warranted.

\subsection{Al$^{37}$Cl}
We have analysed the observed strength of the 
Al$^{37}$Cl  $ J = 17 - 16$ line using an 
excitation model similar to that used to interpret 
observations of formaldehyde and water 
vapor \citep{mel01}.   For the case of AlCl, 
our analysis is hindered by a lack of 
reliable molecular data.  As far as we are aware, rate 
coefficients have never been 
computed for collisional excitation of AlCl, and the 
dipole moment \citep[1--2 Debye, 
according to measurements of the Stark 
effect by][]{lid65} remains uncertain by a 
factor of 2. Accordingly, we simply used rate coefficients 
for excitation of CO by H$_2$ 
in place of those for excitation of AlCl, and assumed a 
dipole moment of 1.5 Debye in 
computing the spontaneous radiative decay rates.  
Fortunately, the predicted line fluxes 
are very insensitive to the assumed rate coefficients 
for collisional excitation; they are, 
however, roughly proportional to the square of the 
assumed dipole moment.  Predicted line fluxes
have been corrected assuming a frequency dependent telescope beamsize
of $HPBW=10\arcsec$ at $240\,$GHz.

A best fit to the available measurements of 
millimeter-wave line emission from 
Al$^{37}$Cl  and Al$^{35}$Cl  is obtained for 
a total assumed AlCl abundance of $4 \times 
10^{-8}$ relative to H$_2$ and an assumed 
Al$^{35}$Cl/Al$^{37}$Cl abundance ratio of 3.
In Figure \ref{alcl}, we show the integrated line fluxes 
observed for Al$^{35}$Cl and Al$^{37}$Cl 
as function of $J_u$, the rotational quantum 
number of the upper state.  Open symbols 
apply to Al$^{35}$Cl and filled symbols to Al$^{37}$Cl.  
Squares show the line strengths 
reported in the AlCl discovery paper of \citet{cer87} 
and the 2mm line 
survey of Cernicharo et al.\ (2000; results 
for the $J = 12 - 11$ line of Al$^{37}$Cl are 
excluded because that line is severely blended); 
and the filled circle shows the Al$^{37}$Cl  
$ J = 17 - 16$ flux reported here.  The dashed and 
solid curves, respectively, show the model predictions 
for Al$^{35}$Cl and Al$^{37}$Cl, obtained for 
a total assumed AlCl abundance of 
$4 \times 10^{-8}$ relative to H$_2$ and an assumed 
Al$^{35}$Cl/Al$^{37}$Cl abundance 
ratio of 3.

Figure \ref{alcl} demonstrates that an acceptable 
fit can be obtained for Al$^{37}$Cl  $ J = 17 - 
16$ line that is broadly consistent with the lower-frequency 
transitions detected previously, 
lending support to our line identification.  
The inferred total AlCl abundance of $4 \times 
10^{-8}$ relative to H$_2$ amounts to 8\% of 
the solar abundance of Cl \citep{AG89} 
and 0.6\% of the  solar abundance of Al.  
Because the derived AlCl
abundance is inversely proportional to the 
square of the assumed dipole moment, it is
uncertain by at least a factor of 2. 
Our value for the AlCl abundance is in broad agreement with 
that found by \citet{hig03}, as well as
with the photospheric AlCl abundances predicted 
by models of cool stellar atmospheres 
\citep[e.g.][]{tsu73}.

\section{Conclusions}
We have clearly detected formaldehyde around the carbon-rich
AGB star IRC+10216, and demonstrated that formaldehyde
has an extended source in the circumstellar envelope around
IRC+10216.  The formaldehyde around IRC+10216 is present
at a level of around $1\%$, relative to water vapor, consistent
with formaldehyde abundances in Solar System comets.  Our
observations, combined with 
previous work by \citet{mel01} and \citet{for03} 
lead us to conclude that formaldehyde is a
molecule produced by the photodissociation of an unknown
parent molecule, and that the parent molecule is produced
by vaporization from the surfaces of cometary nuclei orbiting
IRC+10216 in a Kuiper Belt analog.
We did not detect methanol around IRC+10216, despite 
our deep search, and our upper limits
on the methanol abundance ($<0.077\%$ relative to water vapor)
imply that any cometary system around IRC+10216 contains 
significantly less methanol than typical Solar System comets.

Chemical characterization
of extrasolar cometary systems is important for
a variety of reasons.  The chemical composition of cometary
systems can tell us something about the conditions in
protoplanetary disks around forming stars, and therefore
can provide us with interesting information on the process of star 
formation.  The study of extrasolar cometary systems
also bears directly on the emerging field of astrobiology \citep{EC00,OML90}.
In our own Solar System, it is believed that much of
the volatile carbon present on the young Earth was delivered
by comets \citep{CS92}.
Comets contributed volatile carbon both in the form
of simple molecules like formaldehyde and methanol and as
more complex molecules like amino 
acids \citep{and89,chy90,PC99}.
It has been suggested \citep{CC93,OL97} that this extraterrestrial 
volatile carbon contribution
formed the basis of the pre-biotic carbon 
chemistry which led to the formation of extremely
complex molecules and eventually to single-celled organisms
and all life on Earth.
Thus, the chemistry of extrasolar
cometary systems may have important consequences for the
probability of life forming on otherwise habitable 
exoplanets.
Ultimately, we would like to survey the chemical composition
of a large number of extrasolar cometary systems and compare
their compositions to the chemical composition of Solar System 
comets.

\acknowledgments
We are extremely grateful to Silvia Leurini for taking most of our
followup observations.  We would also like to thank Axel Weiss and Clemens
Thum for their help with the service observing program and for giving
us the opportunity to work with an experimental receiver setup.
Thanks as well to Sergio Martin for his 
help at the telescope.
D.A.N. gratefully acknowledges the support of NASA grants NAG5-13114 from
the Long Term Space Astrophysics Program and NAS5-30702 from the SWAS
program.
K. E. S. F. was partially supported by an American 
Dissertation Fellowship from the American
Association of University Women (AAUW).

\clearpage

\begin{deluxetable}{l l}
\tablecaption{Oxygen-bearing Molecules in Comets \label{oxymolecules}}
\tablehead{\colhead{Molecule} 
& \colhead{Abundance Range} 
}
\tablewidth{0pt}

\startdata
H$_2$O & $100.$ \\
CO & $5.-80.$ \\
CO$_2$ & $3.-30.$ \\
CH$_3$OH & $1.-7.$ \\
H$_2$CO & $0.05-4.$ \\
OCS & $0.3$ \\
SO$_2$ & $\le0.001$ \\
\enddata
\tablecomments{Abundances of oxygen-bearing molecules 
detected in Solar System comet comae, listed as a percentage,
relative to water.  From \citet{boc96}.}
\end{deluxetable}

\begin{deluxetable}{l c r}
\tablecaption{Integration Times \label{inttime}}
\tablehead{\colhead{Position} 
& \colhead{Frequency Band (GHz)}
& \colhead{Time (minutes)} 
}
\tablewidth{0pt}

\startdata
center &  73 &  134 \\
center & 141 & 1166 \\
center & 150 &  976 \\
center & 211 &  312 \\
center & 218 &  232 \\
center & 226 &  436 \\
center & 242 & 2452 \\
off-center,  30\arcdeg & 141 & 128 \\
off-center,  75\arcdeg & 141 & 124 \\
off-center, 120\arcdeg & 141 & 128 \\
off-center, 165\arcdeg & 141 & 124 \\
off-center, 210\arcdeg & 141 & 128 \\
off-center, 255\arcdeg & 141 & 122 \\
off-center, 300\arcdeg & 141 & 128 \\
off-center, 345\arcdeg & 141 & 120 \\
off-center,  30\arcdeg & 150 & 328 \\
off-center,  75\arcdeg & 150 & 124 \\
off-center, 120\arcdeg & 150 & 332 \\
off-center, 165\arcdeg & 150 & 124 \\
off-center, 210\arcdeg & 150 & 328 \\
off-center, 255\arcdeg & 150 & 122 \\
off-center, 300\arcdeg & 150 & 328 \\
off-center, 345\arcdeg & 150 & 120 \\
off-center, avg. & 141 & 1002 \\
off-center, avg. & 150 & 1726 \\
off-center, avg. & 242 & 2636 \\
\enddata
\tablecomments{Times listed are on-source plus off-source.}
\end{deluxetable}

\begin{deluxetable}{l c c c c }
\rotate
\tablecaption{Line Fit Parameters \label{linefit}}
\tablehead{\colhead{Species} 
& \colhead{$\nu_{rest}$} 
& \colhead{$\nu_{obs}$}
& \colhead{Line Strength}
& \colhead{$v_{exp}$}\\
\colhead{and Transition}
& \colhead{(MHz)}
& \colhead{(MHz)}
& \colhead{(mK km s$^{-1}$)}
& \colhead{(km s$^{-1}$)}
}
\tablewidth{0pt}

\startdata
HC$_3$N $J=8-7$                                                & 72783.8220\tablenotemark{b}  & 72783.8 $\pm$ 0.5   & 67138 $\pm$ 82  & 14.0 $\pm$ 1.0 \\
H$_2$CO $1_{01}-0_{00}$                                        & 72837.9480\tablenotemark{b}  & \nodata             & $<246,3\sigma$ & \nodata        \\
C$_5$H $^2\Pi_{1/2}\>J=59/2-57/2\,a$\tablenotemark{\dagger}    & 140824.640\tablenotemark{a}  & fixed               & 294 $\pm$ 19    & 14.5 (fixed)   \\
C$_5$H $^2\Pi_{1/2}\>J=59/2-57/2\,b$\tablenotemark{\dagger}    & 140833.469\tablenotemark{a}  & fixed               & 294 $\pm$ 19    & 14.5 (fixed)   \\
H$_2$CO $2_{12}-1_{11}$\tablenotemark{\dagger}                 & 140839.5020\tablenotemark{b} & fixed               & 687 $\pm$ 19    & 14.5 (fixed)   \\
NaCN $9_{37}-8_{36}$\tablenotemark{\dagger}                    & 140853.858\tablenotemark{c}  & fixed               & 704 $\pm$ 19    & 14.5 (fixed)   \\
SiC$_2$ $6_{25}-5_{24}$                                        & 140920.1478\tablenotemark{c} & 140920.3 $\pm$ 0.5  & 34751 $\pm$ 19  & 14.6 $\pm$ 0.2 \\
NaCN $9_{36}-8_{35}$\tablenotemark{\dagger}                    & 140937.428\tablenotemark{c}  & fixed               & 679 $\pm$ 19    & 14.5 (fixed)   \\
$^{30}$SiC$_2$ $6_{24}-5_{23}$                                 & 140955.918\tablenotemark{c}  & 140955.4 $\pm$ 0.5  & 1383 $\pm$ 25   & 14.9 $\pm$ 0.2 \\
U150\tablenotemark{\ddagger}                                   & \nodata                      & 150385.9 $\pm$ 0.5  & 1040 $\pm$ 21   & 14.3 $\pm$ 0.2 \\
c-C$_3$H$_2$ $2_{20}-1_{11}$                                   & 150436.5448\tablenotemark{b} & 150436.5 $\pm$ 0.5  & 1049 $\pm$ 21   & 14.2 $\pm$ 0.2 \\
H$_2$CO $2_{11}-1_{10}$                                        & 150498.3340\tablenotemark{b} & 150498.6 $\pm$ 0.5  & 396 $\pm$ 21    & 13.9 $\pm$ 0.2 \\
c-C$_3$H $5_{15}-4_{14} (11/2-9/2) 6-5$\tablenotemark{\dagger} & 211117.5764\tablenotemark{b} & 211117.6 $\pm$ 0.5  & 1132 $\pm$ 17   & 14.7 $\pm$ 0.3 \\
c-C$_3$H $5_{15}-4_{14} (11/2-9/2) 5-4$\tablenotemark{\dagger} & 211117.8341\tablenotemark{b} & \nodata             &  \nodata        &  \nodata      \\
H$_2$CO $3_{13}-2_{12}$                                        & 211211.4680\tablenotemark{b} & 211211.5 $\pm$ 0.5  & 652 $\pm$ 17    & 14.6 $\pm$ 0.3 \\
U218c\tablenotemark{\dagger}\tablenotemark{\ddagger}           & \nodata                      & 218085.8 $\pm$ 0.75 & 1734 $\pm$ 31   &  14.5 (fixed)  \\
U218b\tablenotemark{\dagger}                                   & \nodata                      & 218103.1 $\pm$ 0.75 & 353 $\pm$ 31    & 14.1 $\pm$ 0.4 \\
c-C$_3$H$_2$ $5_{24}-4_{13}$                                   & 218160.4423\tablenotemark{b} & 218160.2 $\pm$ 0.75 & 625 $\pm$ 31    & 15.0 $\pm$ 0.4 \\
H$_2$CO $3_{03}-2_{02}$                                        & 218222.1920\tablenotemark{b} & 218222.0 $\pm$ 0.75 & 450 $\pm$ 31    & 14.3 $\pm$ 0.4 \\
U218a                                                          & \nodata                      & 218287.3 $\pm$ 0.75 & 3180 $\pm$ 31   & 14.4 $\pm$ 0.4 \\
HC$_3$N $J=24-23$                                              & 218324.7880\tablenotemark{b} & 218324.7 $\pm$ 0.75 & 13976 $\pm$ 31  & 14.2 $\pm$ 0.4 \\
H$_2$CO $3_{12}-2_{11}$                                        & 225697.7750\tablenotemark{b} & 225697.8 $\pm$ 0.5  & 598 $\pm$ 15    & 14.2 $\pm$ 0.3 \\
Al$^{37}$Cl $J=17-16$                                          & 241855.0339\tablenotemark{b} & 241855.06 $\pm$ 0.3 & 1440 $\pm$ 16   & 14.0 $\pm$ 0.1 \\
\enddata
\tablenotetext{\dagger}{Line is blended.}
\tablenotetext{\ddagger}{Line includes bad channels or other anomalies.}
\tablecomments{We list the line fit parameters for spectra displayed
in Figures \ref{cspec} and \ref{folspec}, and the 
$^{30}$SiC$_2$ $6_{24}-5_{23}$ line.  
The $^{30}$SiC$_2$ line was near the edge of our band at our first
local oscillator setting and but was out of the band
at our second local oscillator setting, thus, it received a bit
more than half the integration time that other lines
in that band received (resulting in somewhat
higher noise levels), and it does not appear in the final
spectrum displayed in Figure \ref{cspec}.
Observed frequencies have been corrected for source motion, 
assuming a systemic velocity of
$v_{LSR}=-26.5{\rm \,km\,s^{-1}}$.
It is important to note that the errors listed in this table 
represent only the statistical errors ($1\sigma$)
associated with each measurement; in several cases, systematic
errors are more important than the statistical errors, and we
estimate that the line strengths for lines with blends or
bad channels may
be uncertain by up to $50\%$.}
\tablenotetext{a}{\citet{gue03}}
\tablenotetext{b}{\citet{pic98}}
\tablenotetext{c}{\citet{CGK00}}
\end{deluxetable}

\begin{deluxetable}{r c c c c }
\tablecaption{Off-Center Integrated Line Strengths in 
${\rm mK\, km\, s^{-1}}$ as a Function of Position Angle 
\label{offline}}
\tablehead{\colhead{Position} 
& \colhead{H$_2$CO} 
& \colhead{c-C$_3$H$_2$}
& \colhead{SiC$_2$}
& \colhead{H$_2$CO} \\
\colhead{Angle}
& \colhead{$2_{11}-1_{10}$}
& \colhead{$2_{20}-1_{11}$}
& \colhead{$6_{25}-5_{24}$}
& \colhead{$2_{12}-1_{11}$}
}
\tablewidth{0pt}

\startdata
30$^{\circ}$  & 153 $\pm$ 25 & 631 $\pm$ 25 & 18681 $\pm$ 27 & 371 $\pm$ 27 \\
75$^{\circ}$  & 118 $\pm$ 88 & 779 $\pm$ 88 & 19661 $\pm$ 40 & 503 $\pm$ 40 \\
120$^{\circ}$ & 138 $\pm$ 23 & 688 $\pm$ 23 & 18664 $\pm$ 30 & 522 $\pm$ 30 \\
165$^{\circ}$ & 80  $\pm$ 55 & 678 $\pm$ 55 & 14449 $\pm$ 41 & 476 $\pm$ 41 \\
210$^{\circ}$ & 216 $\pm$ 23 & 626 $\pm$ 23 & 14795 $\pm$ 27 & 515 $\pm$ 27 \\
255$^{\circ}$ & 77  $\pm$ 53 & 788 $\pm$ 53 & 14882 $\pm$ 43 & 482 $\pm$ 43 \\
300$^{\circ}$ & 124 $\pm$ 25 & 676 $\pm$ 25 & 14211 $\pm$ 31 & 311 $\pm$ 31 \\
345$^{\circ}$ & 165 $\pm$ 56 & 756 $\pm$ 56 & 19737 $\pm$ 36 & 566 $\pm$ 36 \\
\enddata
\tablecomments{We list the line strengths for spectra displayed
in Figures \ref{8aspec} and \ref{8bspec}.  All line strengths
are in units of mK km s$^{-1}$ and errors are $1\sigma$.  Data for the
H$_2$CO $2_{11}-1_{10}$, c-C$_3$H$_2$ $2_{20}-1_{11}$ and
SiC$_2$ $6_{25}-5_{24}$ transitions are also plotted (in
normalized units) in Figure \ref{position}.}
\end{deluxetable}

\begin{deluxetable}{l c c c c }
\tablecaption{Off-Center Integrated Line 
Strengths Relative to Central Position Integrated
Line Strengths \label{offlratio}}
\tablehead{\colhead{Species} 
& \colhead{$\nu_{obs}$} 
& \colhead{$(\int T_A^* dv)_{off}$}
& \colhead{$v_{exp}$}
& \colhead{$(\int T_A^* dv)_{off}/$} \\
\colhead{and Transition}
& \colhead{(MHz)}
& \colhead{(mK km s$^{-1}$)}
& \colhead{(${\rm km\,s^{-1}}$)}
& \colhead{$(\int T_A^* dv)_{cent}$}
}
\tablewidth{0pt}

\startdata
C$_5$H $^2\Pi_{1/2}\>J=59/2-57/2\,a$\tablenotemark{\dagger}& 140824.6 $\pm$ fixed & 233 $\pm$ 13 & 13.6 $\pm$ fixed & 0.79 $\pm$ 0.07 \\
C$_5$H $^2\Pi_{1/2}\>J=59/2-57/2\,b$\tablenotemark{\dagger}& 140833.5 $\pm$ fixed & 233 $\pm$ 13 & 13.6 $\pm$ fixed & 0.79 $\pm$ 0.07 \\
H$_2$CO $2_{12}-1_{11}$\tablenotemark{\dagger}         & 140839.5 $\pm$ fixed & 353 $\pm$ 13 & 13.6 $\pm$ fixed  & 0.51 $\pm$ 0.02 \\
NaCN $9_{37}-8_{36}$\tablenotemark{\dagger}& 140853.9 $\pm$ fixed & 203 $\pm$ 13 & 13.6 $\pm$ fixed & 0.29 $\pm$ 0.02 \\
SiC$_2$ $6_{25}-5_{24}$                    & 140920.0 $\pm$ 0.01 & 17684 $\pm$ 13 & 13.6 $\pm$ 0.02 & 0.5089 $\pm$ 0.0005 \\
NaCN $9_{36}-8_{35}$\tablenotemark{\dagger}& 140937.4 $\pm$ fixed & 177 $\pm$ 13 & 13.6 $\pm$ fixed & 0.26 $\pm$ 0.02 \\
$^{30}$SiC$_2$ $6_{24}-5_{23}$             & 140955.0 $\pm$ 1.0   & 678 $\pm$ 13 & 13.8 $\pm$ 2.0   & 0.49 $\pm$ 0.01 \\
U150\tablenotemark{\ddagger}               & 150385.5 $\pm$ 0.3   & 473 $\pm$ 10 & 12.5 $\pm$ 0.5   & 0.46 $\pm$ 0.01 \\
c-C$_3$H$_2$ $2_{20}-1_{11}$               & 150436.2 $\pm$ 0.1   & 698 $\pm$ 10 & 12.9 $\pm$ 0.2   & 0.67 $\pm$ 0.02 \\
H$_2$CO $2_{11}-1_{10}$                    & 150497.9 $\pm$ 1.0   & 157 $\pm$ 10 & 13.5 $\pm$ 1.2   & 0.40 $\pm$ 0.03 \\
Al$^{37}$Cl $J=17-16$                      & 241854.0 $\pm$ 0.2   & 116 $\pm$ 12 & 13.4 $\pm$ 0.3   & 0.081 $\pm$ 0.008 \\
\enddata
\tablenotetext{\dagger}{Line is blended.}
\tablenotetext{\ddagger}{Line includes bad channels.}
\tablecomments{We list the line strengths for spectra displayed
in Figure \ref{offspec}.
Observed frequencies have been corrected
for source motion, assuming
a systemic velocity of
$v_{LSR}=-26.5{\rm \,km\,s^{-1}}$.  As in Table \ref{linefit}, we list
only statistical errors ($1\sigma$); as in our central position observations, 
systematic errors may be quite
large for blended lines or those with bad channels.  We also
calculate the ratio $(\int T_A^* dv)_{off}/(\int T_A^* dv)_{cent}$
for each line, based on the central position line strengths listed
in Table \ref{linefit}.  The errors quoted for the ratios are also
only statistical.}
\end{deluxetable}

\begin{deluxetable}{l c}
\tablecaption{Excitation Model Parameters \label{CSEparams}}
\tablehead{\colhead{Physical Conditions} 
& \colhead{Value Adopted} 
}
\tablewidth{0pt}

\startdata
%Stellar radius\tablenotemark{a}                   & $7.65\times 10^{13}\,$cm                                      \\
%Photospheric temperature\tablenotemark{a}         & $2,010\,$K                                                    \\
Dust shell inner radius\tablenotemark{a}, $R_i$   & $2.58\times 10^{14}\,$cm                                      \\
Dust shell outer radius\tablenotemark{a}          & $1\times 10^{17}\,$cm                                         \\
Outflow velocity\tablenotemark{a}, $v_{out}$      & $14.5[1.00-0.95(R_i/r)]^{0.5}{\rm \,km\,s^{-1}}$                            \\
Turbulent velocity\tablenotemark{a}               & $0.65{\rm \,km\,s^{-1}}$                                                    \\
H$_2$ density\tablenotemark{b,\dagger}   & $(3.11\times 10^7\,{\rm cm^{-3}}/R_{15}^2)\times(\dot M/3\times 10^{-5})\times(14.5/v_{out})$ \\
Gas temperature\tablenotemark{b,\ddagger} & Max[$10,12(90/R_{15})^{0.72}$]$\,$K                           \\
%Dust temperature                 & $1300(r/R_i)^{-0.4}\,$K                                      \\
%Opacity due to dust              & $0.5\,$cm$^2 (\lambda/50\,\mu{\rm m})^{-1.3}$ per gram of gas \\
Assumed distance\tablenotemark{a}, $D$            & $170\,$pc                                                     \\
\enddata
\tablenotetext{\dagger}{Here $R_{15}$ is the radial 
distance (in cm) divided by $10^{15}$,
$\dot M$ is in units of ${\rm M_{\odot}\,yr^{-1}}$ 
and $v_{out}$ is in units of ${\rm \,km\,s^{-1}}$.}
\tablenotetext{\ddagger}{Max[x,y] means that this quantity
assumes the larger of the two values x and y.}
\tablenotetext{a}{\citet{ski99}}
\tablenotetext{b}{\citet{gla96}}
\end{deluxetable}

\clearpage

\begin{figure}
\plotone{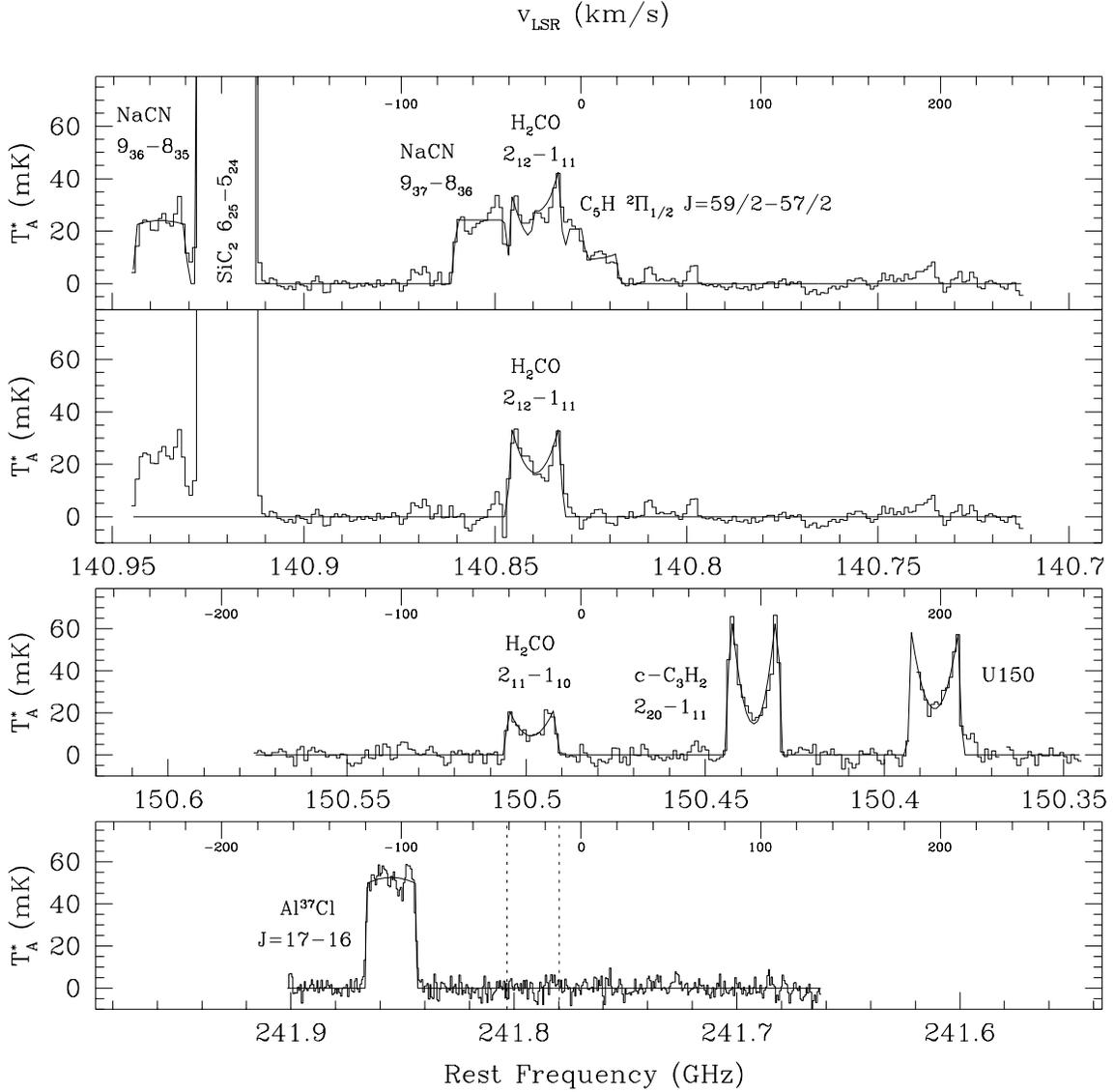}
\caption[Central position spectra showing formaldehyde detection.]
{-- Central position spectra showing formaldehyde detection.
Plotted frequencies represent the frequency in the rest frame
of the source (IRC+10216) assuming a systemic velocity of
$v_{LSR}=-26.5{\rm \,km\,s^{-1}}$.  Frequencies are plotted along the
bottom axis, while $v_{LSR}$ is plotted along the top
axis, with $v_{LSR}=-26.5{\rm \,km\,s^{-1}}$ centered on the frequency
of interest -- $140.840\,$GHz in the top two panels,
$150.498\,$GHz in the next panel, and $241.791\,$GHz in the
bottom panel.  All
lines have been fit using the CLASS {\bf shell} method for lines
observed in circumstellar envelopes.  Line fit parameters
are listed in Table \ref{linefit}.  See text and Table
\ref{linefit} for
(caption continued)
\label{cspec}}
\end{figure}
\addtocounter{figure}{-1}
\begin{figure}
\caption[(continued)]{(continued) further discussion of 
individual line fits.  The
bottom panel displays our spectrum around 
$241.8\,$GHz, which was expected to 
contain a number of
methanol lines.  The extent of the brightest expected
line (centered at $241.791\,$GHz) is 
marked by vertical dotted lines.  
The sole detected line in this panel is tentatively
identified as the $J=17-16$ transition of 
Al$^{37}$Cl (see text for further discussion on 
line
identification and non-detection of methanol).
The second panel from the bottom displays our spectrum around
$150.498\,$GHz, which contains one line of formaldehyde (the $2_{11}-1_{10}\>$
transition), as well as one line of c-C$_3$H$_2$ and one
unidentified line, labeled U150.  Note that this spectrum, and
the U150 line in particular, contains
several bad channels due to our filterbank.  These appear as blank
regions in this plot.
In the top panel, we see that there
are four lines involved in a blend around $140.840\,$GHz.
The NaCN $9_{37}-8_{36}$ and 
C$_5$H $^2\Pi_{1/2}\>J=59/2-57/2$ lines are cataloged,
therefore, we are able to force fit them and subtract them
out, leaving the spectrum below, which clearly displays the
H$_2$CO $2_{12}-1_{11}\>$ line.  The top two
panels also show the $6_{25}-5_{24}$ line of SiC$_2$ and
the $9_{36}-8_{35}$ line of NaCN.
}
\end{figure}

\begin{figure}
\plotone{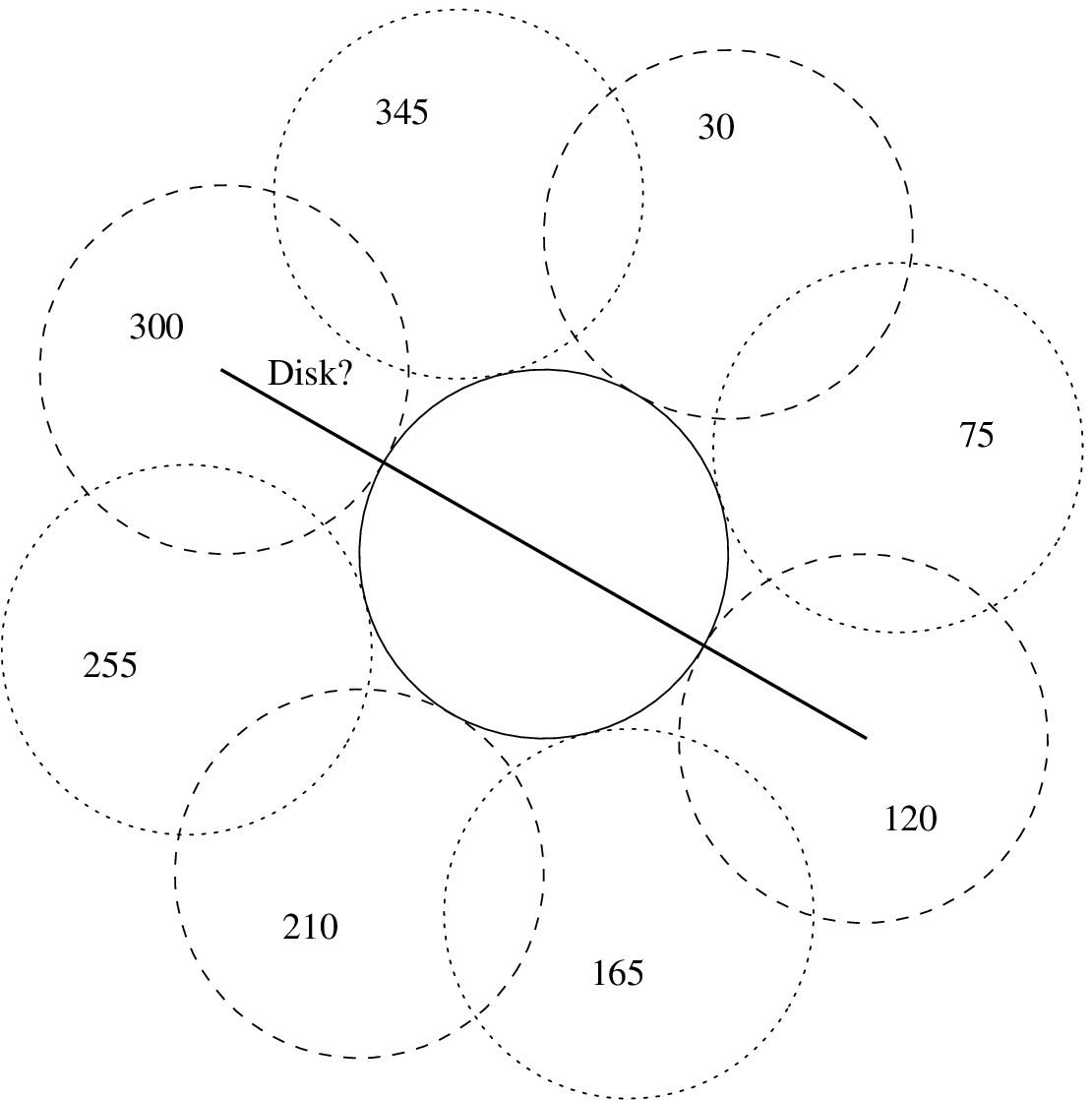}
\caption[Beam pattern used for mapping observations.]
{-- Beam pattern used for mapping observations.  The solid
circle indicates the position of our central
position observations.  The dashed circles
indicate the relative positions of our first set of offset
mapping observations and the dotted circles
indicate the positions of our second set of offset
mapping observations (which received less time than
the first set.  Each circle is labelled with the
appropriate position angle.  The heavy black line,
labeled ``Disk?'' indicates the position angle of a
disk observed by \citet{tut00}
around IRC+10216 in the infrared.
\label{pattern}}
\end{figure}

\begin{figure}
\plotone{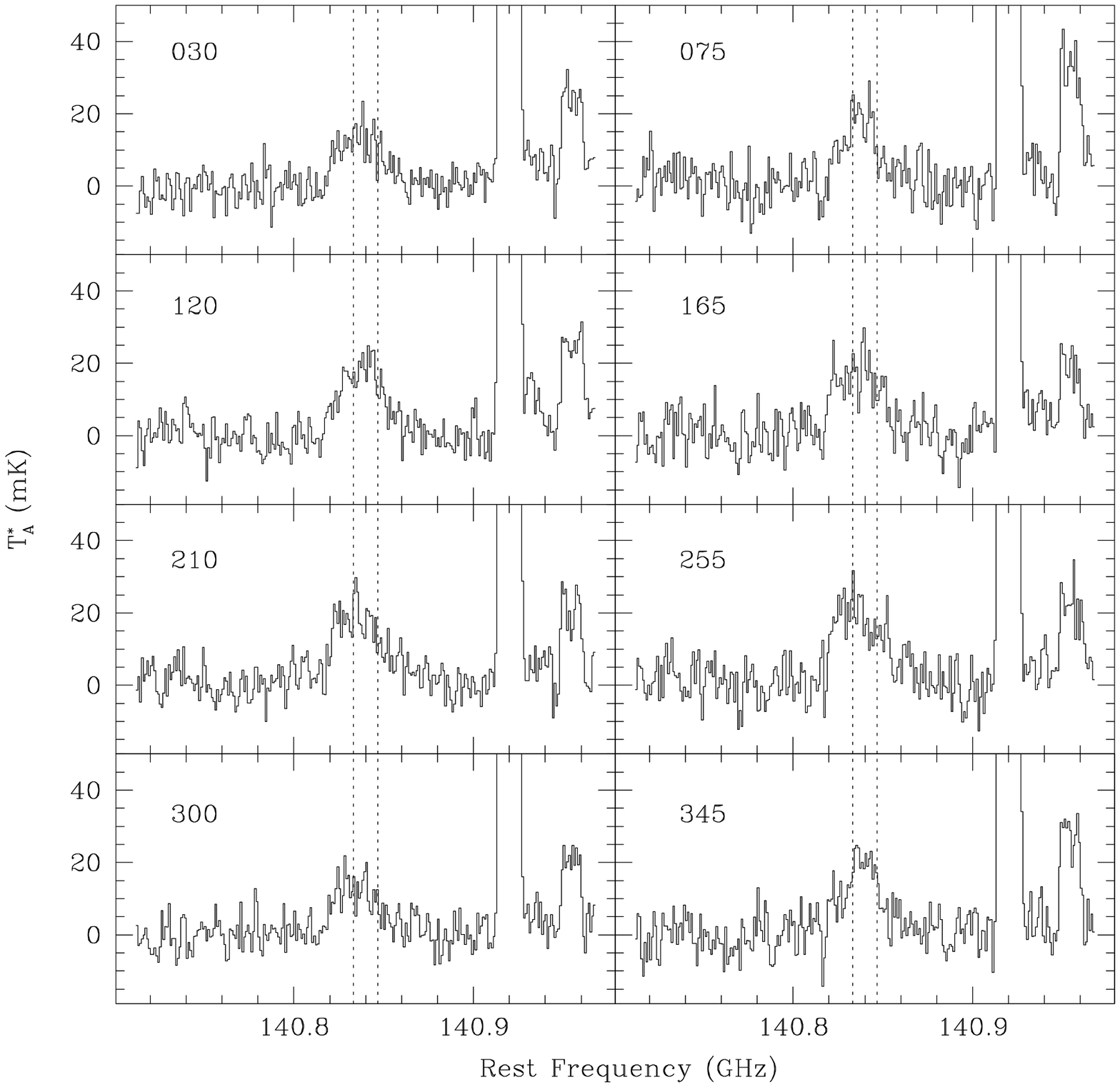}
\caption[Spectra at individual off positions around $140.840\,$GHz.]
{-- Spectra at individual off positions around $140.840\,$GHz.  These
spectra were taken at the labeled position angles (see Figure 
\ref{pattern} for a visual depiction) at a $17\arcsec$ offset
from our central position observations.  The vertical dotted lines
indicate the region of the spectrum where we expect to find
the $2_{12}-1_{11}\>$ transition of H$_2$CO.
\label{8aspec}}
\end{figure}

\begin{figure}
\plotone{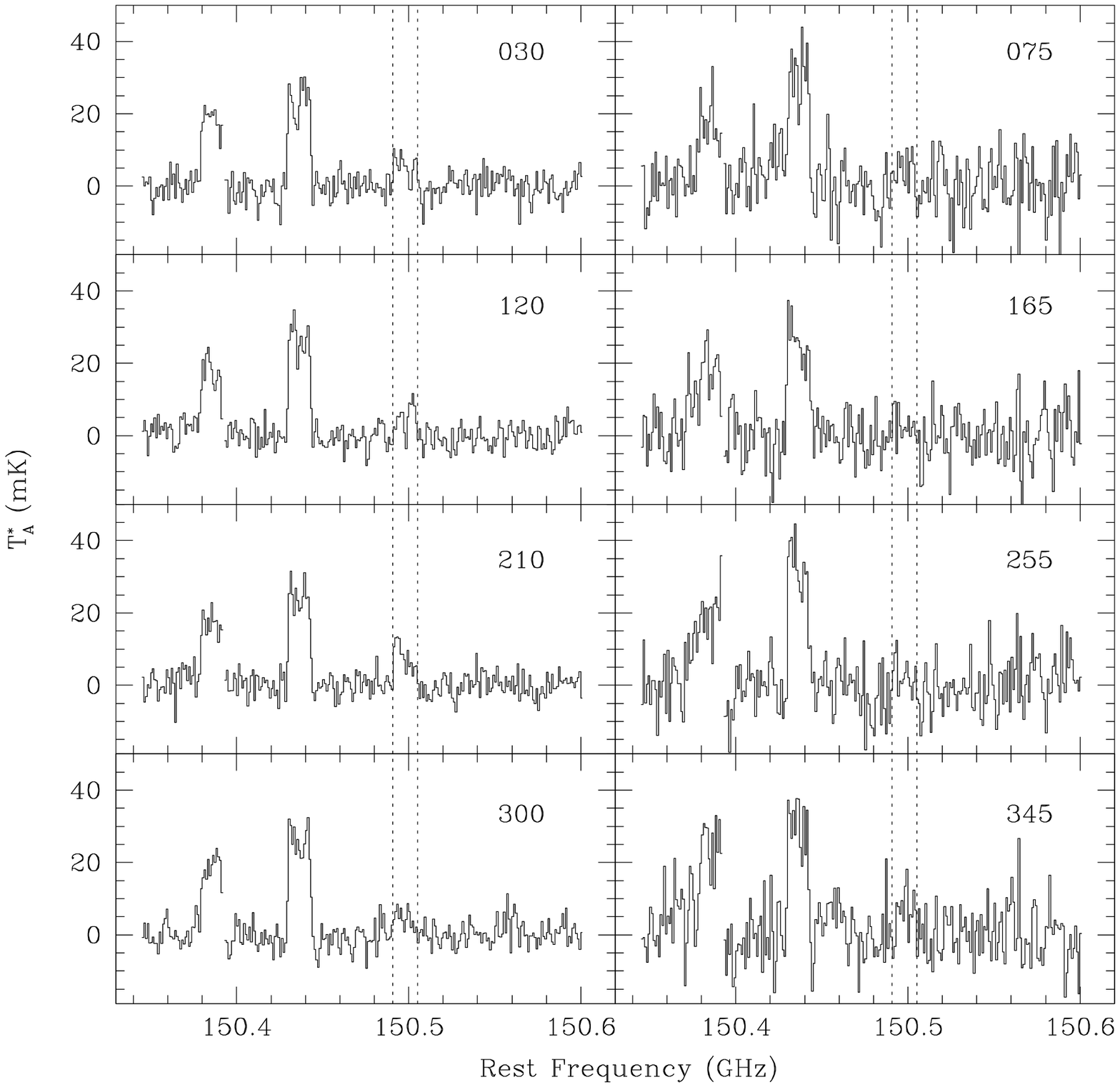}
\caption[Spectra at individual off positions around $150.498\,$GHz.]
{-- Spectra at individual off positions around $150.498\,$GHz.  These
spectra were taken at the labeled position angles (see Figure 
\ref{pattern} for a visual depiction) at a $17\arcsec$ offset
from our central position observations.  The vertical dotted lines
indicate the region of the spectrum where we expect to find
the $2_{11}-1_{10}\>$ transition of H$_2$CO.
\label{8bspec}}
\end{figure}

\begin{figure}
\plotone{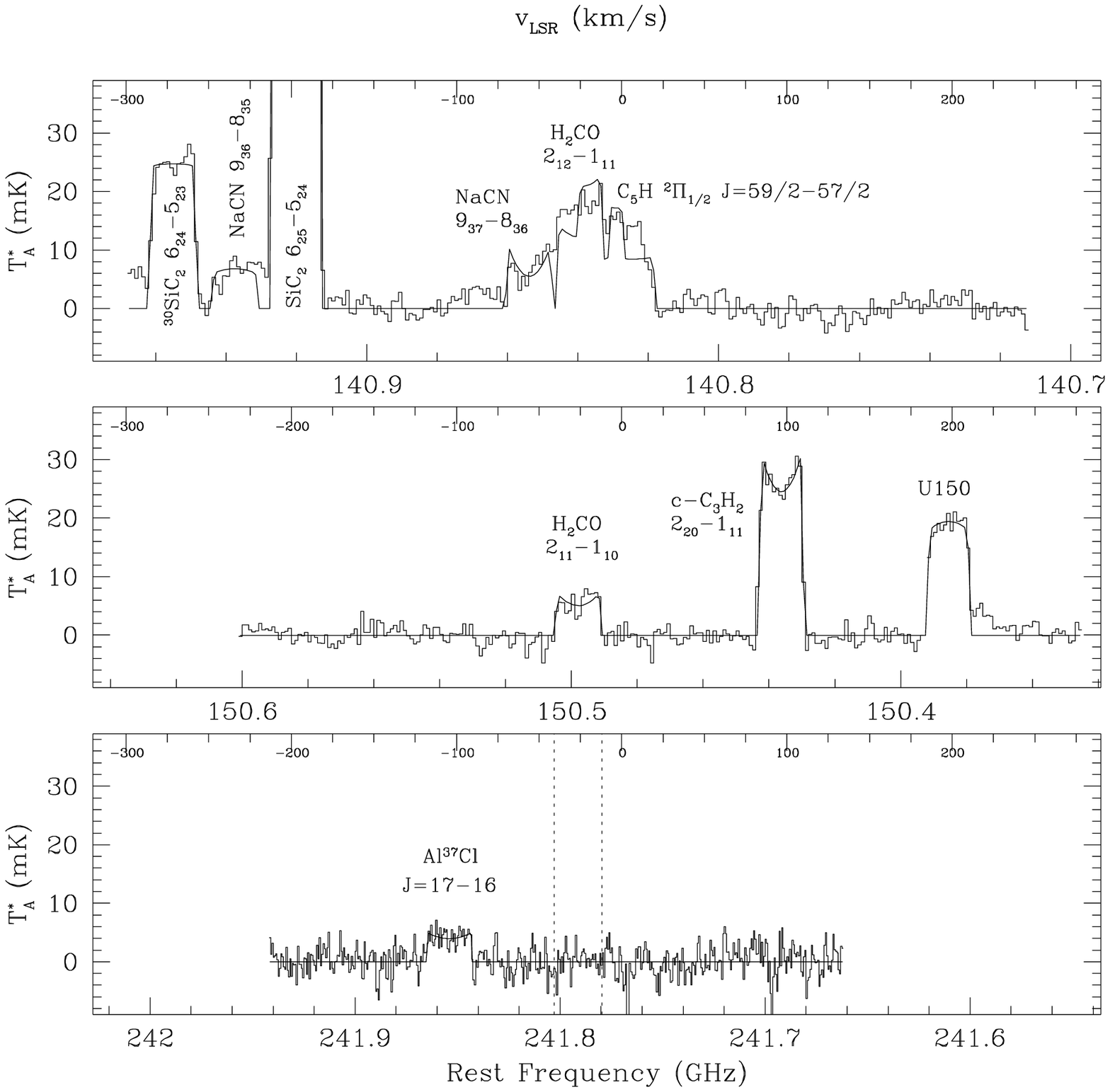}
\caption[Spectra of average off position.]
{-- Spectra of average off position. These spectra were produced
by co-adding the eight individual off position spectra to
create a single average off position.  This plot follows
the conventions of Figure \ref{cspec}, except that
we have not displayed the subtracted residuals from
the $140.840\,$GHz spectrum.  Also note the different
(smaller) temperture scale, as compared with Figure 
\ref{cspec}.  The line parameters for the
displayed fits can be found in Table \ref{offlratio}.
\label{offspec}}
\end{figure}

\begin{figure}
\plotone{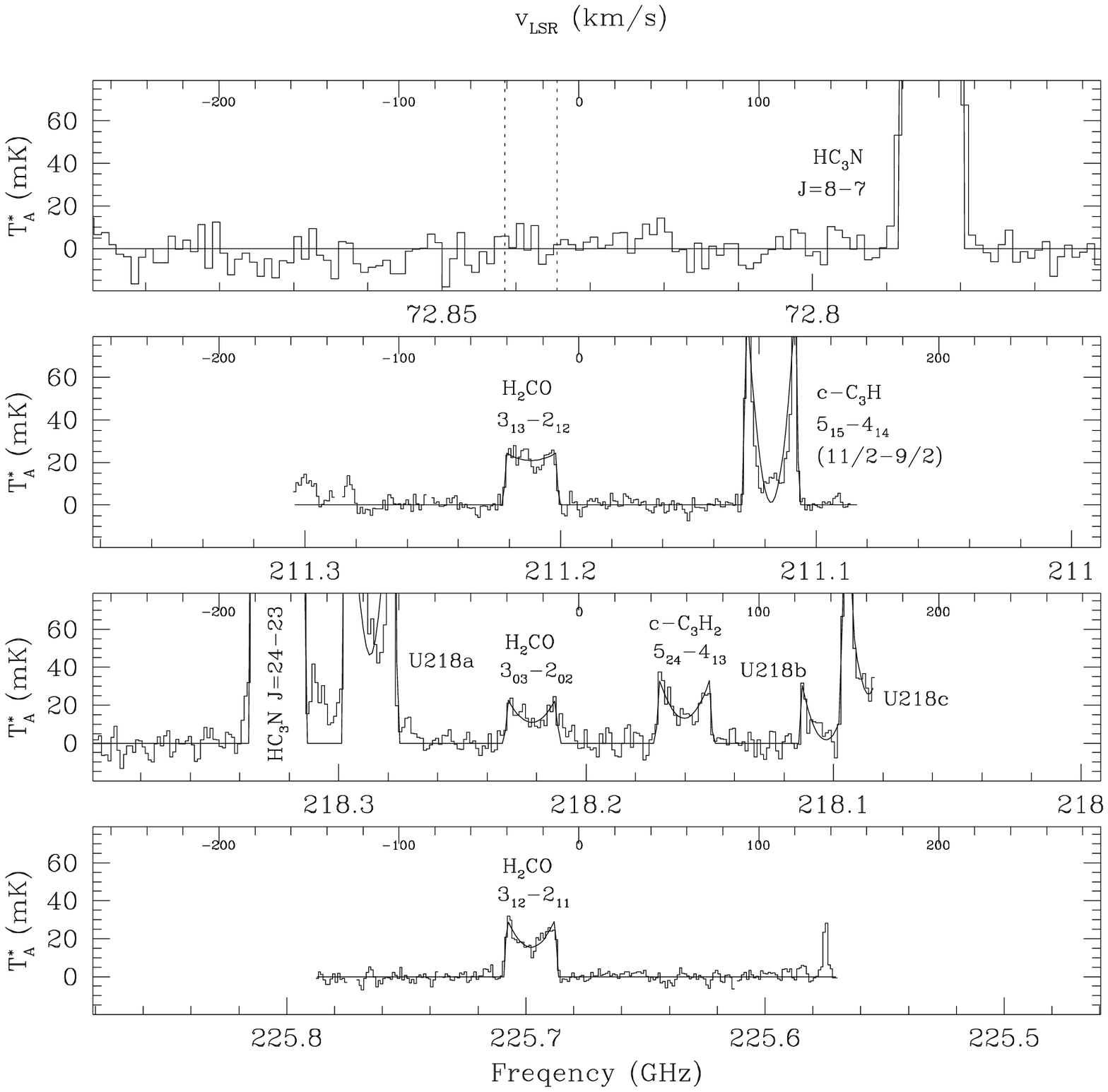}
\caption[Followup spectra at central position.]
{-- Followup spectra at central position.  This plot follows
the conventions of Figure \ref{cspec}.  The line parameters for the
displayed fits can be found in Table \ref{linefit}.  The dotted lines
in the top panel indicate the expected position of the $1_{01}-0_{00}$
transition of formaldehyde (see text for further details).
\label{folspec}}
\end{figure}

\begin{figure}
\plotone{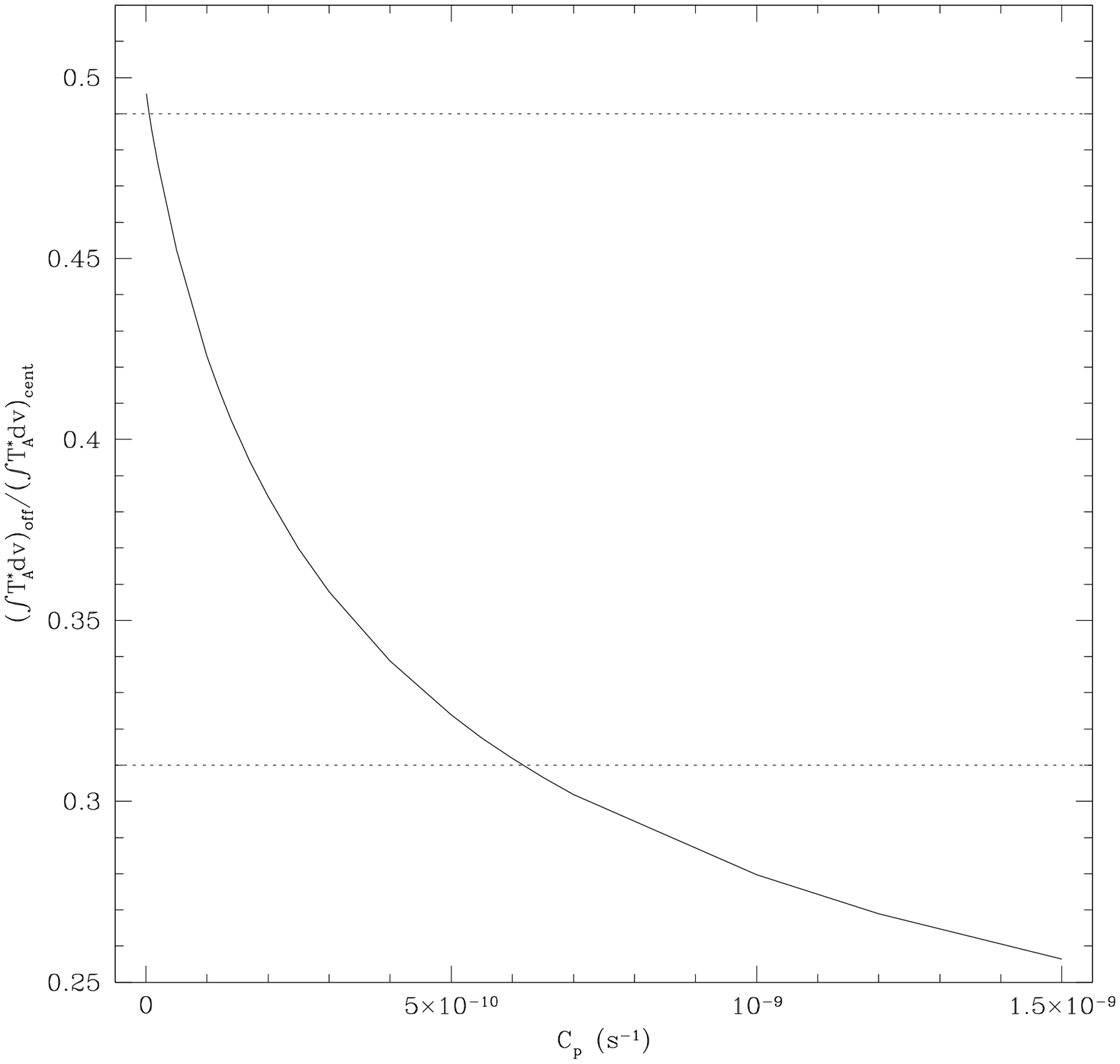}
\caption[Photodissociation rate of formaldehyde parent vs. ratio
of off position flux to central position flux.]
{-- Photodissociation rate of formaldehyde parent vs. ratio
of off position flux to central position flux.  We plot the
predicted ratio of off position to central position flux
($(\int T_A^* dv)_{off}/(\int T_A^* dv)_{cent}$) for
formaldehyde lines, for a given $C_p$, the unshielded
photodissociation rate of formaldehyde's parent molecule.
The dotted horizontal lines denote the $3\sigma$ limits for the
ratio measured from the $2_{11}-1_{10}\>$ transition of
formaldehyde.  Due to the large uncertainty 
in the measured ratio, $C_p$ is not well constrained,
and any value of $C_p$ between about $1\times 10^{-11}$
and $7\times 10^{-10}$ will produce a ratio consistent
with our observations.
\label{ratio}}
\end{figure}

\begin{figure}
\plotone{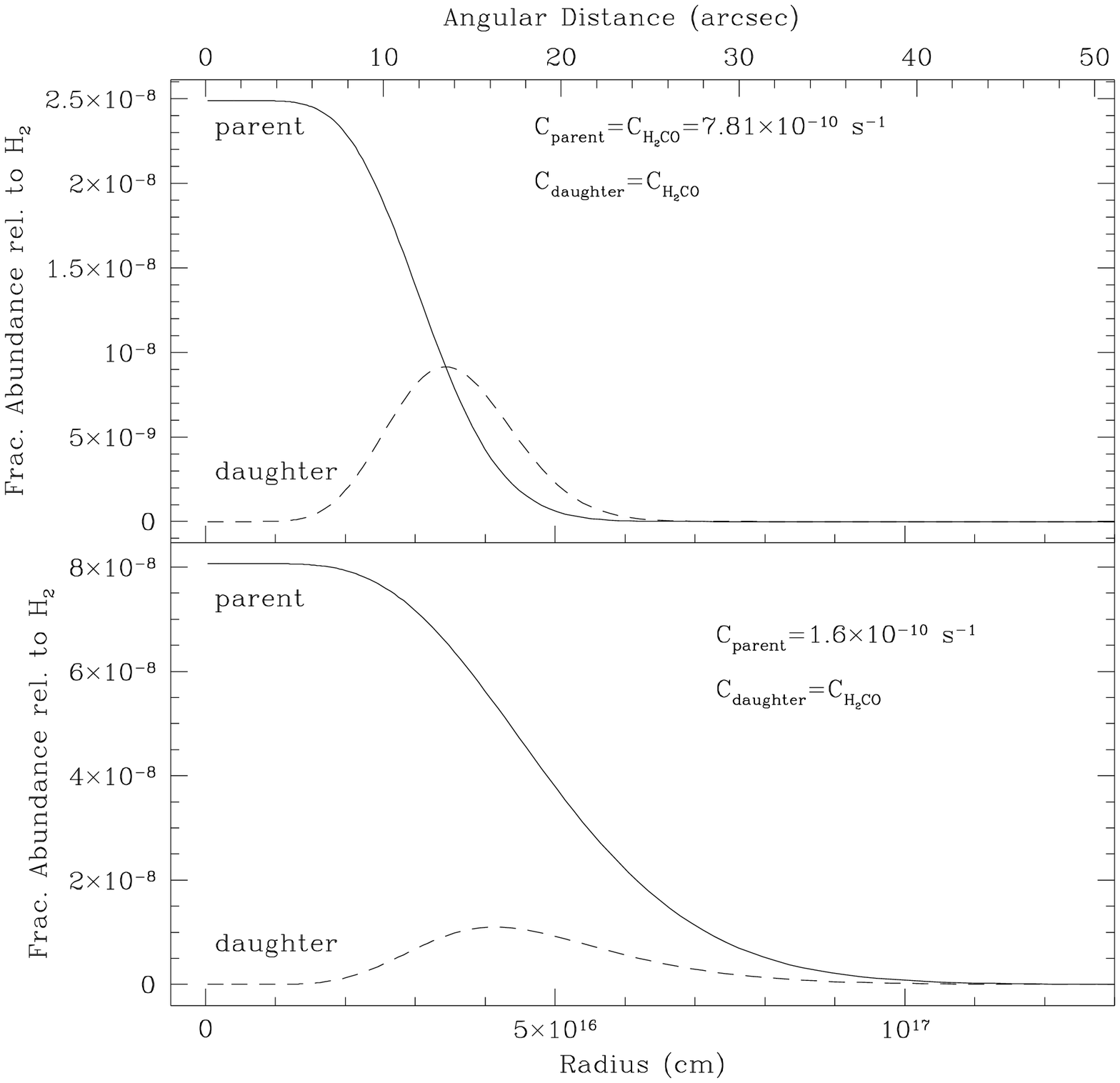}
\caption[Radial distribution of formaldehyde and formaldehyde parent.]
{-- Radial distribution of formaldehyde and formaldehyde parent.  
We plot the radial distributions a parent molecule (solid line)
which is photodissociated into a daughter molecule (dashed line).
The angular distance across the top axis is computed for an
assumed stellar distance of $170\,$pc.
In the top panel, the unshielded photodissociation rate for
both molecules is the same as the unshielded rate for formaldehyde.
Thus, the solid line in the top panel shows the radial distribution
of formaldehyde if it were a parent molecule.  The dashed line
in the top panel shows the formaldehyde distribution if it is a
daughter product whose parent molecule has an identical
photodissociation rate to that of formaldehyde.  
In the lower panel, we display our
(caption continued)
\label{photdiss}}
\end{figure}
\addtocounter{figure}{-1}
\begin{figure}
\caption[(continued)]{(continued) best fit model for 
the spatial distribution of formaldehyde.
In this model, formaldehyde is a daughter product, and the
unshielded photodissociation rate of formaldehyde's parent
molecule is $C_p=1.7\times10^{-10}\,$s$^{-1}$.  Both panels
have abundances normalized so that the daughter product
produces a signal at
$150.498\,$GHz (the $2_{11}-1_{10}\>$ transition of formaldehyde) 
in the central beam of our observations that matches
our observed line strength.}
\end{figure}

\begin{figure}
\plotone{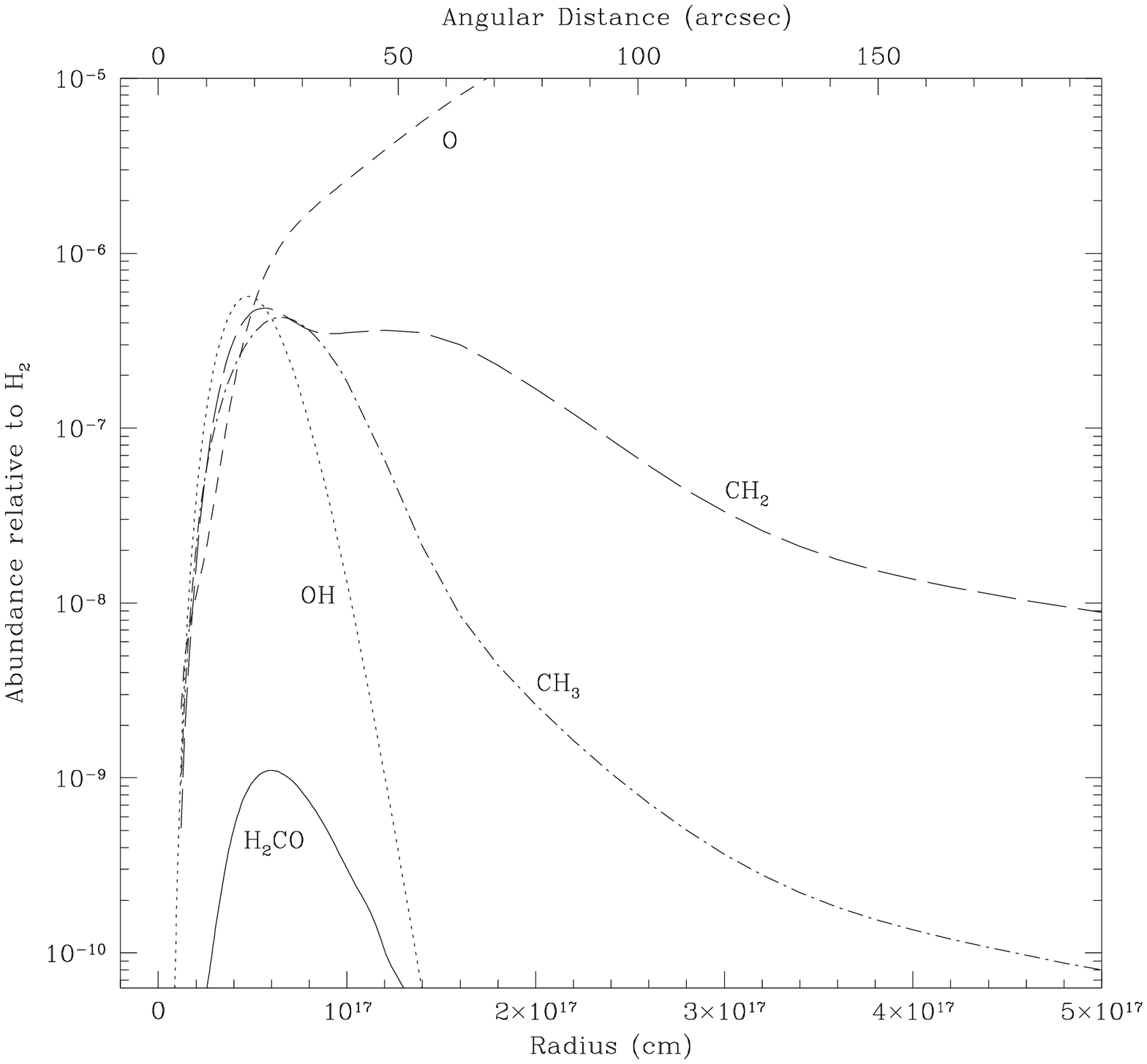}
\caption[Chemical production of formaldehyde in IRC+10216.]
{-- Chemical production of formaldehyde in IRC+10216.  Formaldehyde
could be produced chemically in IRC+10216 through reactions
of O with CH$_3$ and OH with CH$_2$; however our models
indicate that only relatively small amounts of formaldehyde will actually
be produced, given the slow reaction rates, large interstellar
UV field and 
low densities involved.  We plot the 
abundance relative to
molecular hydrogen of the various molecules, 
assuming only chemical production of formaldehyde, 
as a function of radius.  The angular distance 
across the top axis is computed for an
assumed stellar distance of $170\,$pc.
Note that the peak abundance of formaldehyde
predicted by the chemical models is an order of magnitude
smaller than the observed abundance.
\label{chem}}
\end{figure}

\begin{figure}
\plotone{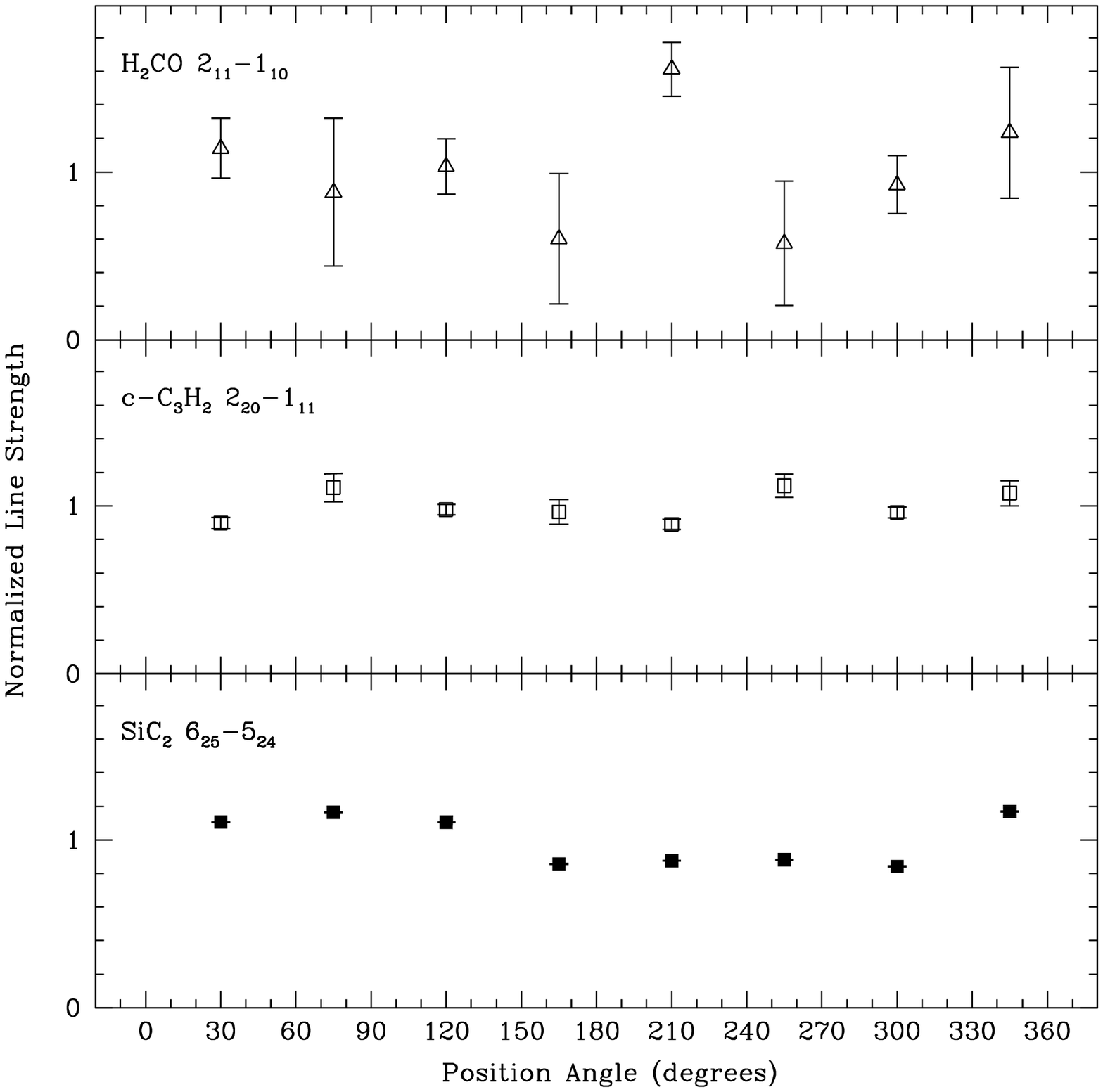}
\caption[Line strength versus position angle.]
{-- Line strength versus position angle.  We show the
variation of line strength with position angle (see
Figure \ref{pattern} for mapping grid) for three lines:
H$_2$CO $2_{11}-1_{10}$, c-C$_3$H$_2$ $2_{20}-1_{11}$ and 
SiC$_2$ $6_{25}-5_{24}$.  The quantity plotted on the
y-axis is normalized line strength, computed by dividing
the measured line strength at each position angle by the average
of the line strengths at all eight positions.  
The c-C$_3$H$_2$ and SiC$_2$
lines come from molecules produced chemically in 
the circumstellar
envelope around IRC+10216, so we regard them as
``control lines'' -- that is, they indicate the normal
variation with position angle of the circumstellar
envelope. (caption continued)
\label{position}}
\end{figure}
\addtocounter{figure}{-1}
\begin{figure}
\caption[(continued)]{(continued)
By contrast, formaldehyde is believed
to be present around IRC+10216 because of a vaporizing
Kuiper Belt analog, and so might be expected to have
a different spatial distribution from the molecules
that produce our control lines.  One point,
at $210^{\circ}$, does depart substantially from
the average line strength of formaldehyde, and this
departure is not seen in the control lines.  However,
the line strength of formaldehyde does not appear to 
vary in a
periodic fashion with a period of 
$180^{\circ}$, as would
be expected for an edge-on disk or ring.  Still, at this
signal to noise level it is difficult to rule out
a ring with any certainty.}
\end{figure}

\begin{figure}
\epsscale{0.70}
\plotone{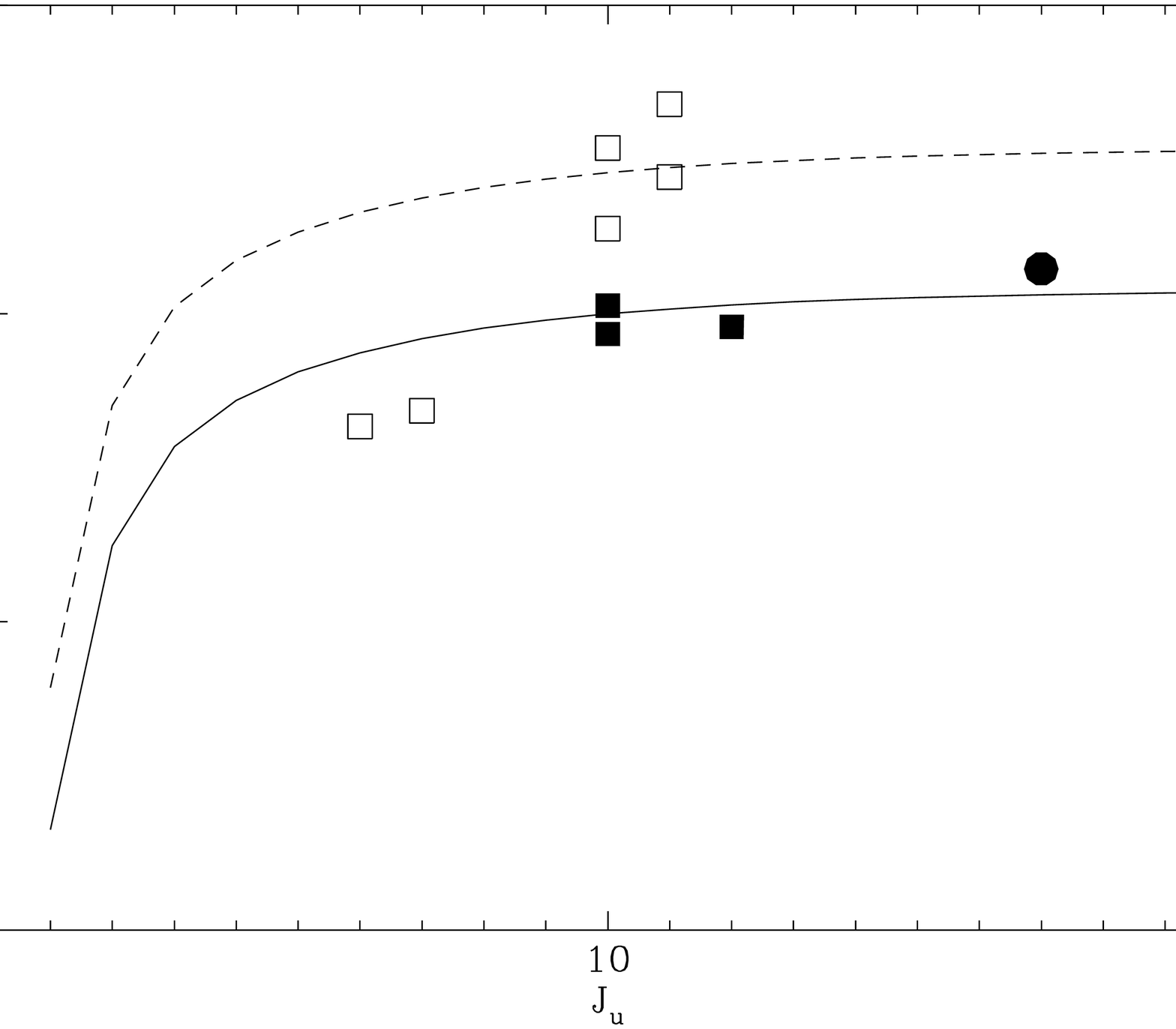}
\caption
{-- Integrated line fluxes observed for Al$^{35}$Cl and Al$^{37}$Cl 
as function of $J_u$, the rotational quantum number of 
the upper state.  Open symbols 
apply to Al$^{35}$Cl and filled symbols to Al$^{37}$Cl.  
Squares show the line strengths 
reported in the AlCl discovery paper of \citet{cer87} 
and the 2mm line 
survey of Cernicharo et al.\ (2000; results 
for the $J = 12 - 11$ line of Al$^{37}$Cl are 
excluded because that line is severely blended); 
and the filled circle shows the Al$^{37}$Cl  
$ J = 17 - 16$ flux reported here.  The dashed and 
solid curves, respectively, show the model predictions 
for Al$^{35}$Cl and Al$^{37}$Cl, obtained for 
a total assumed AlCl abundance of 
$4 \times 10^{-8}$ relative to H$_2$ and an assumed 
Al$^{35}$Cl/Al$^{37}$Cl abundance 
ratio of 3.
\label{alcl}}
\end{figure}

\end{document}